# On the Implementation of Two-Parameter Stabilizing Controllers Using Stable-Block Input-Output Feedback Structures

**Fernando di Sciascio***

**Instituto de Automática UNSJ-CONICET, Av. San Martín (Oeste) 1109, San Juan, J5400ARL, Argentina**

*Abstract. This paper a universal implementation method for continuous- and discrete-time two-parameter (2-DOF) stabilizing controllers using exclusively stable constituent blocks. The framework is valid for all regular SISO and MIMO plants, particularly those failing to satisfy the parity interlacing property where unstable controllers are unavoidable. Unlike existing algebraic approaches that compromise parametric design freedom to force blocks stability, the proposed method preserves design precedence. By leveraging the Youla-Kučera parametrization within input-output (I/O) feedback topologies, we decouple synthesis from implementation. This allows physical performance specifications to drive the controller design, which is subsequently realized via a cyclic configuration of stable blocks. This architecture yields key engineering advantages, including localized anti-windup, simplified initialization, and improved numerical robustness.*

## I. INTRODUCTION

In control engineering, ensuring internal stability is a basic prerequisite for any controller design method [1], [2]. A linear feedback control system is internally stable if none of its components contain hidden unstable modes and the injection of bounded external signals at any place in the system results in bounded output signals measured anywhere in the system [2, Definition 4.4, p. 134]. This means that every possible transfer function from any input to any output must be BIBO stable. The problem of parametrizing, for a given plant, the set of all internally stabilizing single-loop (one-parameter, or one-degree-of-freedom, 1-DOF) controllers was completely solved by Youla, Bongiorno, and Jabr in the landmark paper [3], and independently formulated by V. Kučera using an algebraic approach [4]. Since then, the Youla-Kučera (YK) parametrization of all stabilizing controllers has become a cornerstone in feedback control of linear systems. Subsequently, Vidyasagar extended the YK parametrization of all stabilizing controllers to two-parameter (or two-degree-of-freedom, 2-DOF) controllers [5].

In practice, except for controllers with deliberate integral action, the use of unstable controllers is avoided unless strictly necessary, due to several technical drawbacks, including: (i) fragility under loop failure [2]; (ii) startup and reset challenges [6]; (iii) vulnerability to actuator saturation[7], [8], [9]; (iv) drastic reduction of the phase margin and conditional stability [2]; and (v) implementation sensitivity to numerical errors [10]. The notion of stable or strong stabilization was introduced by Youla, Bongiorno, and Lu [11];

* Corresponding author.
E-mail address: fernando@inaut.unsj.edu.ar

the strong stabilization problem entails designing a stable feedback controller that ensures the internal stability of a given plant. A necessary and sufficient condition for the existence of a stable stabilizing controller is that the plant satisfies the parity interlacing property (PIP); i.e., the plant has an even number of poles between any pair of its zeros on the extended positive real axis.

Clearly, this result implies that strong stabilization is an intrinsic property of the plant, as it depends solely on the locations of its real positive poles and zeros. This enables the definition of the class of strongly stabilizable Single-Input Single-Output (SISO) or Multi-Input Multi-Output (MIMO) plants. The controller architecture—whether one- or two-parameter—has no bearing on strong stability. This point is clarified by the following quote from M. Vidyasagar's influential book [5, p. 149]: "Problems such as strong stabilization, …, are no different in the two-parameter case from the one-parameter case. This is because the feedback compensator has to stabilize the plant in order to achieve stability, irrespective of which scheme is used."

When an open-loop unstable plant fails to satisfy the PIP, internal stabilization by means of a stable controller is fundamentally impossible. Under these circumstances, an unstable controller can be realized either by incorporating at least one unstable block within an acyclic topology, or by employing exclusively stable blocks within a cyclic topology—i.e., a configuration embedding at least one internal feedback loop. The latter represents the best possible design for an unstable two-parameter controller, as it ensures that the constituent blocks remain stable while strictly preserving the independent degrees of freedom required for performance synthesis.

The engineering advantages of this stable-block architecture are multi-faceted, as it inherits the core operational benefits typically associated with purely stable controllers: (i) independent verification and testing [2]; (ii) facilitated initialization and startup procedures enabling bumpless transfer [5]; (iii) advanced saturation management via localized anti-windup [8], [9]; and (iv) improved numerical conditioning that reduces fragility [10].

In this paper, we derive structural parametrizations of the controller blocks based on the Youla-Kučera framework, specifically tailored for a family of five input-output (I/O) feedback topologies [12], [13, pp. 422–425]. By mapping the stable Youla-Kučera free parameters directly to these architectures, we introduce two-parameter controller implementations where the constituent blocks are inherently stable for both SISO and MIMO systems. Rather than forcing stable controller blocks through ad-hoc transfer functions, the methodology proposed herein preserves the traditional order of design precedence. This approach is motivated by the fact that purely algebraic schemes achieve block stability at the expense of the parametric freedom required for closed-loop synthesis. Consequently, essential properties such as transient performance and closed-loop robustness are relegated to collateral outcomes of the mathematical formalism.

This shortcoming becomes particularly severe when dealing with open-loop unstable plants. In accordance with Stein's seminal paradigm to "respect the unstable" [7], unstable dynamics impose strict analytic limitations that cannot be bypassed by algebraic manipulation. Attempting to force the stability of controller blocks through structural over-constraint does not eliminate the inherent fragility of the unstable plant.

Conversely, the framework presented in this work positions system performance specifications as the central design criterion rather than a secondary outcome. Instead of focusing on specific synthesis techniques, this paper proposes an implementation method for two-parameter stabilizing controllers for regular (proper) SISO and MIMO plants using stable blocks. That is, to address a specific design problem, the designer is free to adopt the most convenient or preferred synthesis method, and subsequently implement the controller with entirely stable components by utilizing the control structures and their respective parametrizations proposed in this work.

Except within numerical examples, for notational convenience no distinction is made between signals and their Laplace transforms. Owing to the abstract algebraic framework adopted in this work, all derived results apply seamlessly to both the continuous-time and discrete-time cases. The synthesis of stabilizing controllers is formulated over a general ring $\mathcal{S}$ of stable and proper rational transfer functions, or the matrix ring $M(\mathcal{S})$ of stable and proper rational transfer matrices. The choice of the time domain merely redefines the geometric region of stability for this ring—mapping poles to the open left-half complex plane for continuous systems, or to the interior of the unit disk ($|z| < 1$) for discrete-time systems.

The rest of this paper is organized as follows. Section II lays the necessary groundwork by revisiting internal stability and coprime factorizations, followed by a treatment of two-parameter stabilizing control laws, their implementation topologies, and input-output equivalences. Section II concludes with an introduction to five controller implementations of the I/O Feedback topology family. Section III presents the main results, detailing the parametrization of all stabilizing two-parameter controllers for both SISO and MIMO systems. Specifically, it covers the five input-output feedback controller implementations featuring a stable-block structure, with several numerical examples provided for validation. Finally, concluding remarks are provided in Section IV.

**Notation:** With a few exceptions, the notation used throughout this paper is standard. For clarity, we summarize the main symbols and conventions below.

$\mathcal{R}$ : Set of proper rational functions.

$\mathcal{S}$ : Set of proper stable rational functions (the standard $\mathcal{RH}_\infty$ is replaced by $\mathcal{S}$ for brevity).

$\mathcal{U}$ : The set of units of the ring $\mathcal{S}$, consisting of all bistable (i.e., biproper, stable, and minimum-phase) rational functions.

$M(\mathcal{R})$ : Set of matrices whose elements belong to $\mathcal{R}$.

$M(\mathcal{R})^{m\times n}$ : Set of $m \times n$ matrices over $\mathcal{R}$.

$M(\mathcal{S})$ : Set of matrices whose elements belong to $\mathcal{S}$.

$M(\mathcal{S})^{m\times n}$ : Set of $m \times n$ matrices over $\mathcal{S}$.

$\mathcal{U}(\mathcal{S})$ : Set of unimodular matrices over $\mathcal{S}$ i.e., $Q \in \mathcal{U}(\mathcal{S}) : Q, Q^{-1} \in M(\mathcal{S})$.

$\mathcal{H}$ : Set of Hurwitz polynomials.

$(\alpha, \beta, \gamma)$ : Implementation of the 2-DOF control law $u = C_{ff} r - C_{fb} y$ using the blocks $\alpha$, $\beta$, $\gamma$.

$C^{(\alpha,\beta,\gamma)}(P)$ : Controller with blocks $\alpha$, $\beta$ and $\gamma$.

$u^{(\alpha,\beta,\gamma)}(P)$ : Control signal (output of controller $C^{(\alpha,\beta,\gamma)}$ and input to plant $P$.

$T_{y_i,u_j}^{(\alpha,\beta,\gamma)}(P)$ : Closed-loop transfer function from input $u_j$ to output $y_i$ for a plant $P$ equipped with a controller $C^{(\alpha,\beta,\gamma)}$.

$\cong$ : Input-output equivalence of two implementations through their respective controllers.

$\deg(p)$ : Degree of a polynomial $p$.

$\rho(P)$ : Relative degree of a transfer function $P = n / d$, i.e., $\rho = \deg(d) - \deg(n)$.

$T_s$ : Discretization sampling period.

## II. THEORETICAL PRELIMINARIES

This section reviews fundamental results regarding Youla-Kučera parameterization, the set of all stabilizing controllers, and two-degree-of-freedom architectures. With the appropriate modifications, these results apply to both discrete- and continuous-time systems. For notational simplicity, signals and their Laplace transforms are not distinguished.

### A. Internal Stability [1], [2]

A linear feedback control system is internally stable if none of its components contain hidden unstable modes and the injection of bounded external signals at any place in the system results in bounded output signals measured anywhere in the system. This means that every possible transfer function from any input to any output must be bounded-input bounded-output (BIBO) stable.

### B. Coprime Factorization [1], [5], [14], [15]

#### i) SISO Case

Every plant $P \in \mathcal{R}$ admits a coprime factorization over $\mathcal{S}$: $P = N / D$ with $N, D \in \mathcal{S}$ that satisfies the Bézout identity $X_1 N + X_2 D = 1$ with $X_1, X_2 \in \mathcal{S}$.

#### ii) MIMO case

Every MIMO plant with $m$ outputs and $n$ inputs $P(s) \in M(\mathcal{R})^{m \times n}$ admits a doubly coprime factorization over $M(\mathcal{S})^{m \times n}$; (left and right) $P = \tilde{D}^{-1}\tilde{N} = ND^{-1}$ with $N, D, \tilde{N}, \tilde{D} \in M(\mathcal{S})$ that satisfies the Bézout matrix identity:

$$\begin{bmatrix} X_2 & X_1 \\ -\tilde{N} & \tilde{D} \end{bmatrix}\begin{bmatrix} D & -\tilde{X}_1 \\ N & \tilde{X}_2 \end{bmatrix} = \begin{bmatrix} I & 0 \\ 0 & I \end{bmatrix}, \quad X_1, X_2, \tilde{X}_1, \tilde{X}_2 \in M(\mathcal{S}), \tag{1}$$

where the dimensions of the constituent matrices are:

$$
\begin{aligned}
&P(s) \in M(\mathcal{R})^{m\times n},\ N,\tilde{N} \in M(\mathcal{S})^{m\times n},\ X_1,\tilde{X}_1 \in M(\mathcal{S})^{n\times m} \\
&D, X_2 \in M(\mathcal{S})^{n\times n},\ \tilde{D},\tilde{X}_2 \in M(\mathcal{S})^{m\times m}
\end{aligned}
\tag{2}
$$

In the doubly coprime factorization $P = \tilde{D}^{-1}\tilde{N} = ND^{-1}$, the pairs $(\tilde{X}_1, \tilde{X}_2)$ and $(X_1, X_2)$ are not independent. If $(\tilde{X}_1, \tilde{X}_2)$ is known, one can obtain $X_1 = \tilde{X}_1 + DW\tilde{N}$ and $X_2 = \tilde{X}_2 - NW\tilde{N}$, where $W \in \mathcal{S}$ is arbitrary. Conversely, if $(X_1, X_2)$ is known, then $\tilde{X}_1$ and $\tilde{X}_2$ can be derived as $\tilde{X}_1 = DX_1\tilde{D}^{-1}$ and $\tilde{X}_2 = (I - \tilde{N}\tilde{X}_1)\tilde{D}^{-1}$.

**C. One- and Two-Parameter Control Laws**

A *control law* is a *causal mapping* that generates the control signal *u* from the plant output signal $y$ and a reference signal $r$. The simplest control structure is the one-parameter control law (3), whose variants are shown in Figs. 1(a) and 1(b).

$$
u = \begin{cases} C(r - y) \\ r - Fy \end{cases} . \tag{3}
$$

Fig. 1. One-parameter control laws: (a) cascade or unity-error feedback, (b) output-feedback or non-unity error feedback.

It is well established that, using the Youla-Kučera parametrization, the set of all stabilizing single-loop controllers can be expressed in terms of a stable free parameter $Q$ [1], [3], [5]:

$$
\mathcal{C}_{stab}(P) = \left\{ C = \frac{X_1 + QD}{X_2 - QN} :\ Q \in \mathcal{S} \right\}, \tag{4}
$$

where $Q$ is an arbitrary stable proper transfer function, subject to the constraint that $X_2 - QN \neq 0$.

In this framework, fundamental design trade-offs must be addressed, such as the balance between robust stability and reference tracking response.

The two-parameter control law in (Fig. 2)

$$
u = C_{ff}r - C_{fb}y, \tag{5}
$$

is the most general control structure involving the external reference $r$ and the plant output $y$. As noted by Vidyasagar [5], a three-parameter compensator does not exist. The pair $(C_{ff}, C_{fb})$ are proper transfer

functions—referred to as the gains of the two-parameter control law—which enable the independent processing of the reference signal and the output measurements [5], [15]. The role of $C_{fb}$—the feedback component—is to ensure internal stability and robustness, whereas $C_{ff}$ is designed to optimize the response to the reference input. This design problem can be formulated using two stable transfer functions, $Q_r$ and $Q_y$, known as free parameters; specifically, $Q_y$ is utilized to provide stability and robustness, while $Q_r$ defines the command (reference signal) response [16], [17].

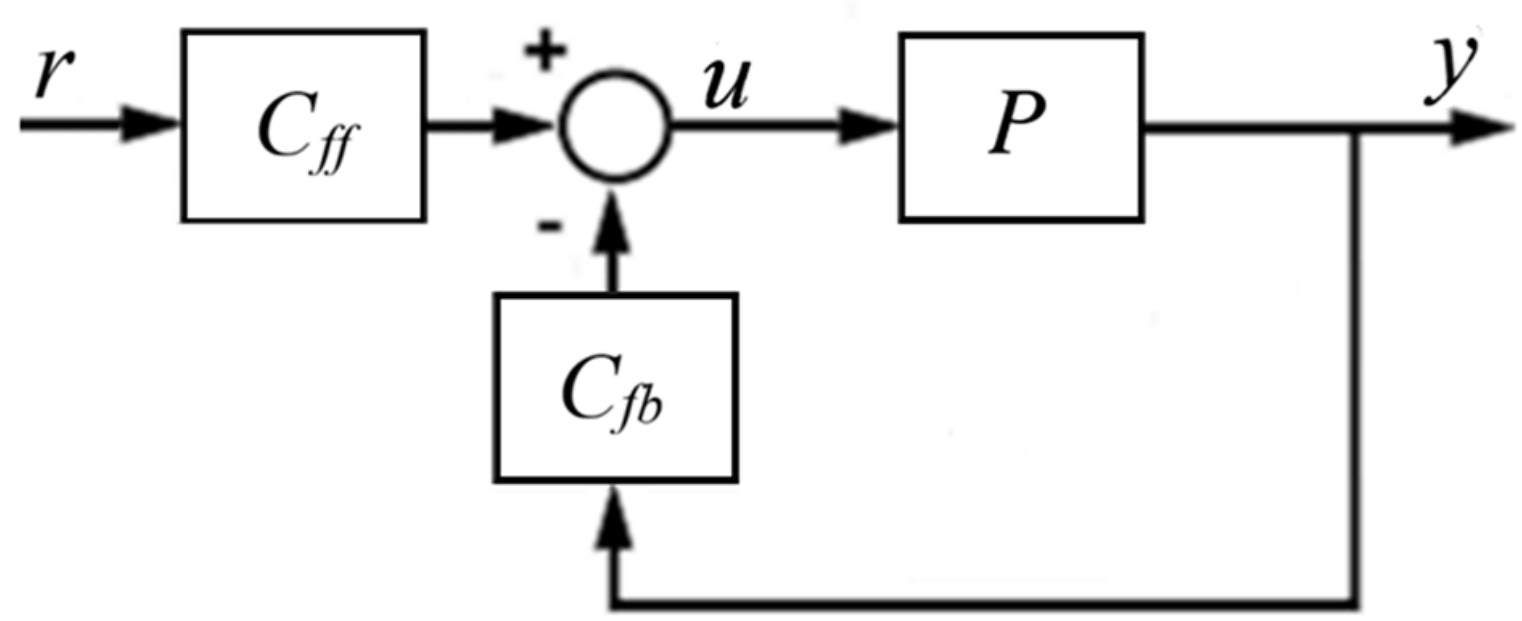

Fig. 2. Implementation (infeasible) $(C_{ff}, C_{fb})$ of Two-Parameter Control Law (5).

**D. Parametrization of all stabilizing two-parameter control laws**

**a) SISO case**

For a given plant $P \in \mathcal{R}$, $\mathcal{C}_{stab}(P)$ denote the set of two-parameter *stabilizing control law gains* $(C_{ff}, C_{fb}) \in \mathcal{R}$, that is,

$$\mathcal{C}_{stab}(P) := \left\{(C_{ff}, C_{fb}) \in \mathcal{R} : \text{the control law } u = C_{ff}r - C_{fb}\,y \text{ is stabilizing}\right\}, \tag{6}$$

and

$\mathcal{S}_{stab}(P)$ denote the set of two-parameter stabilizing controllers.

$$\mathcal{S}_{stab}(P) := \left\{C^{(C_{ff}, C_{fb})} = [C_{ff} \quad -C_{fb}] : (C_{ff}, C_{fb}) \in \mathcal{C}_{stab}(P)\right\}. \tag{7}$$

Using the YK parametrization with $N, D, X_1, X_2 \in \mathcal{S}$, the set $\mathcal{S}_{stab}(P)$ can be parametrized by [5, p. 146], [15, Theorem 10.16. p. 434].

$$\mathcal{C}_{stab}(P) = \left\{\left(C_{ff} = \frac{Q_r}{X_2 - Q_y N}, C_{fb} = \frac{X_1 + Q_y D}{X_2 - Q_y N}\right) : \ Q_r, Q_y \in \mathcal{S}\right\}, \tag{8}$$

where the parameters $Q_r$, $Q_y$ are arbitrary stable proper transfer functions, subject to the constraints that $X_2 - Q_y N \neq 0$ and $Q_r \neq 0$.

It should be noted that the constraint $Q_r \neq 0$ is practical rather than mathematical; $Q_r = 0$ decouples the plant from the reference signal $r$. If the interest is solely in the plant regulation problem, one simply adopts $r = 0$ instead of $Q_r = 0$. Therefore, an arbitrary $Q_r$ implies that $Q_r \in \mathcal{S} \setminus \{0\}$.

In view of (8), a bijective mapping (9) is obtained between parameters $(Q_r, Q_y)$ and $(C_{ff}, C_{fb})$.

$$C_{fb} = \frac{X_1 + Q_y D}{X_2 - Q_y N} \Rightarrow X_2 C_{fb} - X_1 = (D + NC_{fb})Q_y \Rightarrow Q_y = \frac{X_2 C_{fb} - X_1}{D + NC_{fb}}$$

$$C_{ff} = \frac{Q_r}{X_2 - Q_y N} \Rightarrow Q_r = X_2 C_{ff} - NC_{ff} \underbrace{\frac{X_2 C_{fb} - X_1}{D + NC_{fb}}}_{Q_y} = \frac{C_{ff} \overbrace{(NX_1 + X_2 D)}^{1}}{D + NC_{fb}}$$

$$\boxed{\begin{array}{lcl} C_{ff} = \dfrac{Q_r}{X_2 - Q_y N} & & Q_r = \dfrac{C_{ff}}{D + NC_{fb}} \\ & \Leftrightarrow & \\ C_{fb} = \dfrac{X_1 + Q_y D}{X_2 - Q_y N} & & Q_y = \dfrac{X_2 C_{fb} - X_1}{D + NC_{fb}} \end{array}} \tag{9}$$

The two-parameter controller in Fig. 3, with input and output disturbance signals $d$ and $n$ respectively, allows for the analysis of internal stability.

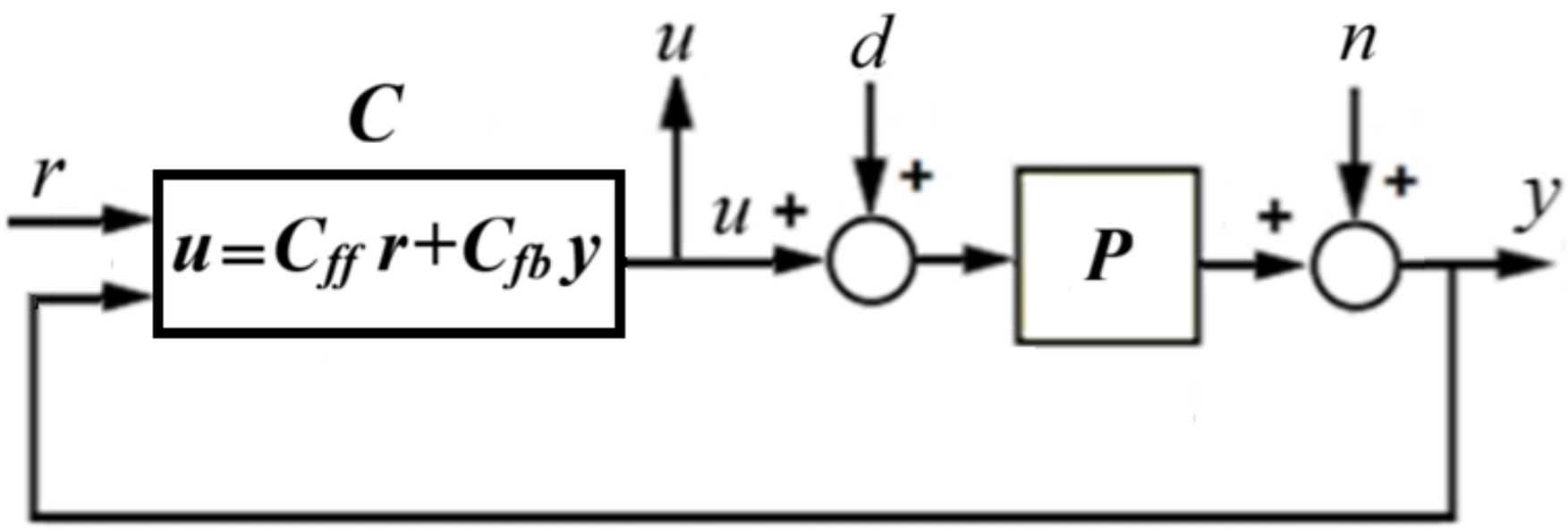

Fig. 3. Two-parameter controller with input and output process disturbances $d$ and $n$, respectively.

Let $H(P, C_{ff}, C_{fb})$ denote the 2×3 transfer matrix from $[r\ d\ n]^{\mathrm{T}}$ to $[y\ u]^{\mathrm{T}}$.

$$\begin{bmatrix} y \\ u \end{bmatrix} = H(P, C_{ff}, C_{fb}) \begin{bmatrix} r \\ d \\ n \end{bmatrix} = \begin{bmatrix} T_{yr} & T_{yd} & T_{yn} \\ T_{ur} & T_{ud} & T_{un} \end{bmatrix} \begin{bmatrix} r \\ d \\ n \end{bmatrix}. \tag{10}$$

According to Section C, the feedback system is internally stable if and only if the six transfer functions in matrix $H(P, C_{ff}, C_{fb})$ are stable.

$$H(P,C_{ff},C_{fb}) = \frac{1}{1+C_{fb}P}\begin{bmatrix} C_{ff}P & P & 1 \\ C_{ff} & -C_{fb}P & -C_{ff} \end{bmatrix}, \tag{11}$$

substituting $P = N / D$ and $C_{ff}$, $C_{fb}$ from (9) into (11) yields:

$$H(N,D,Q_r,Q_y) = \begin{bmatrix} Q_r N & (X_2 - Q_y N)N & (X_2 - Q_y N)D \\ Q_r D & -(X_1 + Q_y D)N & -(X_1 + Q_y D)D \end{bmatrix}. \tag{12}$$

The six transfer functions of matrix $H$ are referred to as the "gang of six" by Åström and Murray [17].

$$\begin{cases} T_{yr} = Q_r N, & T_{yd} = (X_2 - Q_y N)N, & T_{yn} = (X_2 - Q_y N)D \\ T_{ur} = Q_r D, & T_{ud} = -(X_1 + Q_y D)N, & T_{un} = -(X_1 + Q_y D)D \end{cases} \tag{13}$$

The dynamic behavior of the closed-loop system, specifically its response to reference signals, load disturbances, and measurement noise, is governed by (13). All implementations of the two-parameter control law (5) have the same six transfer functions as those in (13).

**b) MIMO case**

For a given plant $P \in M(\mathcal{R})$, the YK parametrization with $X_1, X_2, \tilde{X}_1, \tilde{X}_2 \in M(\mathcal{S})$, the set $\mathcal{S}_{stab}(P)$ can be parametrized by [5, Theorem p. 146], [15, Theorem 10.16. p. 434].

$$\mathcal{C}_{stab}(P) = \quad C_{ff} = (X_2 - Q_y\tilde{N})^{-1}Q_r,\ C_{fb} = (X_2 - Q_y\tilde{N})^{-1}(X_1 + Q_y\tilde{D}) \ : \ Q_r, Q_y \in M(\mathcal{S}) \ , \tag{14}$$

where the parameters $Q_r$, $Q_y$ are arbitrary stable matrix transfer functions, subject to the constraints that $X_2 - Q_y N \neq 0$ and $Q_r \neq 0$.

As previously established for SISO case, the one-to-one correspondence between parameters $Q_r, Q_y \in M(\mathcal{R})^{n\times m}$ and $C_{ff}, C_{fb} \in M(\mathcal{R})^{n\times m}$, and conversely, is straightforward that:

$$\boxed{\begin{matrix} C_{ff} = (X_2 - Q_y\tilde{N})^{-1}Q_r & & Q_r = (X_2 - Q_y\tilde{N})C_{ff} \\ & \Leftrightarrow & \\ C_{fb} = (X_2 - Q_y\tilde{N})^{-1}(X_1 + Q_y\tilde{D}) & & Q_y = (X_2 C_{fb} - X_1)(\tilde{D} + \tilde{N}C_{fb})^{-1} \end{matrix}} \tag{15}$$

Observe that, in both the SISO and MIMO cases, the foregoing parametrization applies, $C_{fb}$ is parametrized by $Q_y$, whereas $C_{ff}$ is parametrized by $Q_r$. The parametrizations (8) and (14) generates the entire set of stabilizing control laws in the sense that each pair $(Q_r, Q_y)$ corresponds to a stabilizing control law (5), while each stabilizing control law corresponds to a pair $(Q_r, Q_y)$.

The feedback controller $C_{fb}$ is designed to meet robust performance requirements in a manner similar as in a single feedback loop (one-degree of freedom controller). The prefilter control $C_{ff}$ is then added to the overall compensated system to optimize the command response independently of the loop robustness and disturbance attenuation governed by $C_{fb}$.

As in the SISO case, the six matrix transfer functions in $H(N, D, \tilde{N}, \tilde{D}, Q_r, Q_y)$ are stable.

$$\begin{cases} T_{yr} = NQ_r, \quad T_{yd} = (\tilde{X}_2 - NQ_y)\tilde{N}, \quad T_{yn} = (\tilde{X}_2 - NQ_y)\tilde{D} \\ T_{ur} = DQ_r, \quad T_{ud} = -(\tilde{X}_1 + DQ_y)\tilde{N}, \quad T_{un} = -(\tilde{X}_1 + DQ_y)\tilde{D} \end{cases} \tag{16}$$

**E. Implementation of a two-parameter stabilizing control law**

A clear distinction is drawn by Kwok and Davison [18] between the two-parameter stabilizing control law (5) and its implementation: the latter refers to the realization of the mathematical expression $u = C_{ff}r - C_{fb}y$ as a practical, functional controller. This entails the controller topology— the quantity of SISO blocks and their interconnection scheme—and its configuration—the specific parametrization, such that the control law (5) is preserved.

An implementation of a two-parameter stabilizing control law for a given plant $P \in \mathcal{R}$ is classified as *feasible* if the following properties hold [13], [18]:

i) All controller blocks are proper (open-loop properness).

ii) The overall feedback system is well posed (closed-loop properness).

iii) All possible transfer function from any input to any output must be BIBO stable (internal stability).

Conversely, if an implementation fails to satisfy these requirements—normally internal stability—it is classified as *infeasible*. An implementation is *universally feasible* if it remains feasible for any plant $P \in \mathcal{R}$.

Figure 4 illustrates the direct implementation of the two-parameter stabilizing control law (5) into a two-block two-parameter controller. This implementation, denoted as the $(C_{ff}, C_{fb})$ implementation, is infeasible in practice, as it may result in an unstable feedforward controller $C_{ff}$.

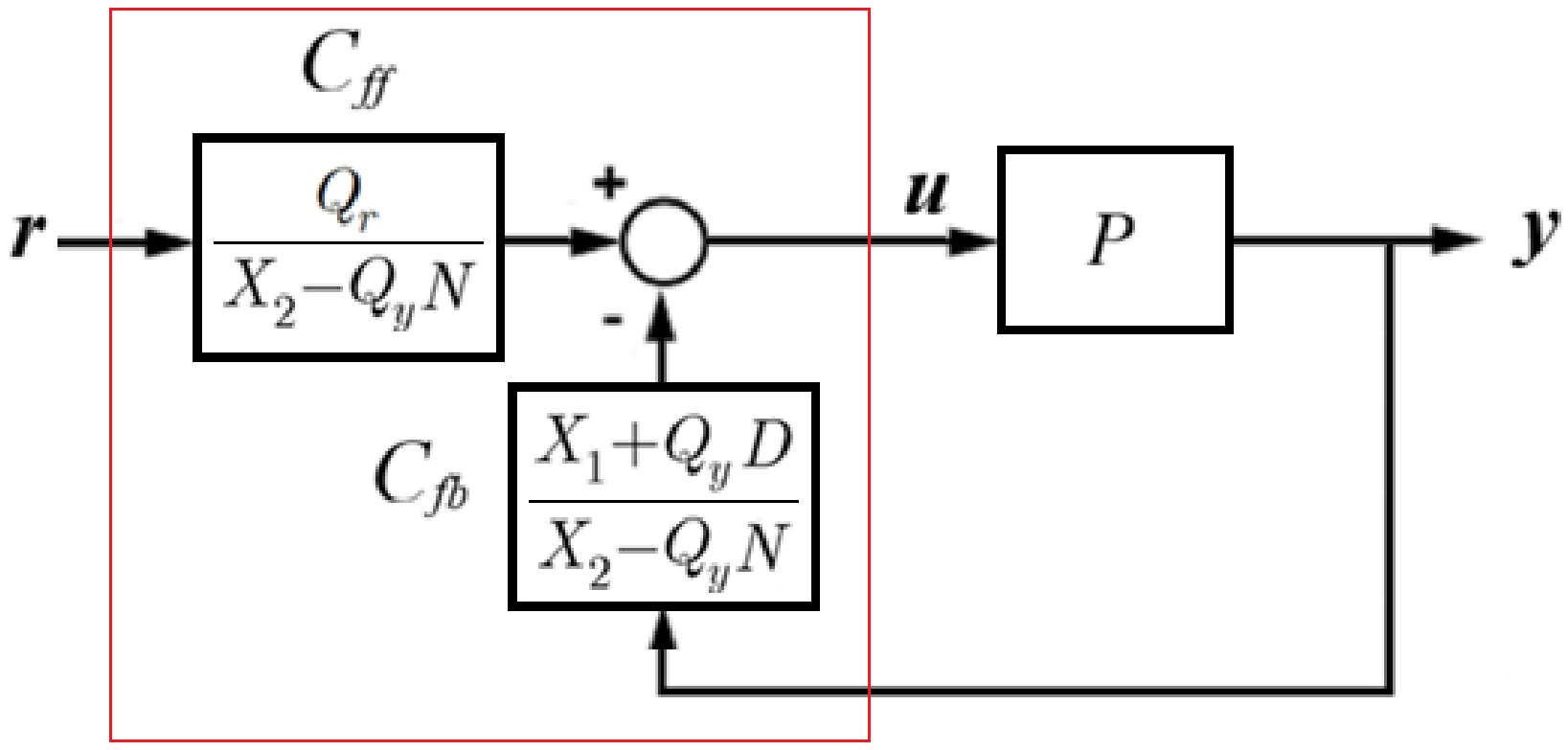

Fig. 4. Two-block Infeasible Implementation of a two-parameter SISO compensator.

As shown in [5] and [15], a universally feasible implementation can be realized through the three-block controller topology depicted in Fig. 4, for both the SISO and MIMO cases, denoted as the $(C_r, C_e, C_y)$ implementation. The set of two-parameter stabilizing controllers can be parametrized as follows.

- SISO case (Fig. 5(a))

$$\mathcal{C}_{stab}(P) = \left\{ \left( C_r = Q_r, C_e = \frac{1}{X_2 - NQ_y}, C_y = X_1 + DQ_y \right), X_2 - NQ_y \neq 0 : \; (Q_r, Q_y) \in \mathcal{S} \right\} \quad (17)$$

- MIMO case (Fig. 5(b)) [15]

$$\mathcal{C}_{stab}(P) = \left\{ \left( C_r = Q_r, C_y = X_1 + Q_y \tilde{D}, C_e = (X_2 - Q_y \tilde{N})^{-1} \right) \text{ with : } \begin{array}{c} \left| X_2 - Q_y \tilde{N} \right| \neq 0 \\ (Q_r, Q_y) \in M(\mathcal{S}) \end{array} \right\} \quad (18)$$

As before, in both the SISO and MIMO cases, the parameters $Q_r$ and $Q_y$ are arbitrary but proper and stable.

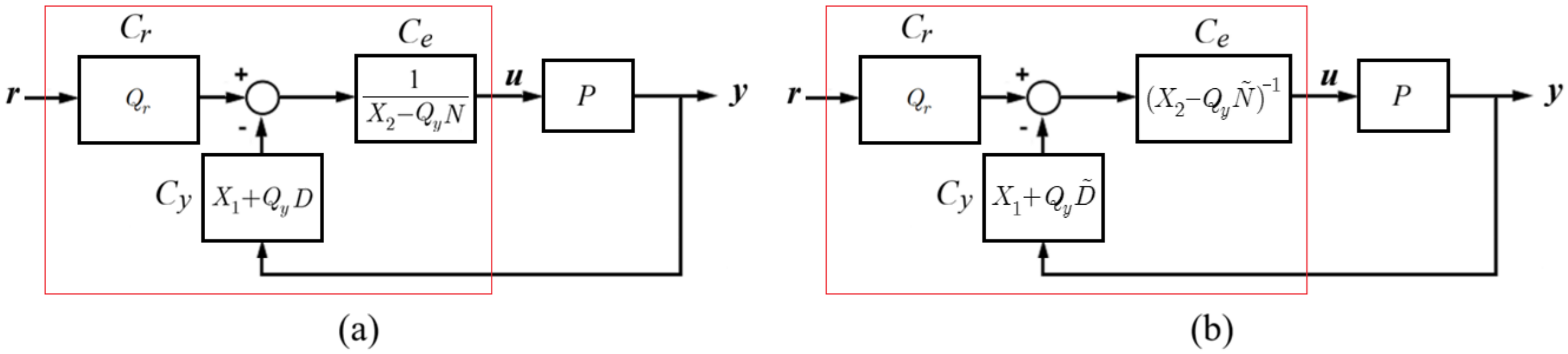

Fig. 5. Standard feasible Implementation of a two-parameter compensator: (a) SISO case, (b) MIMO case.

### F. Input-Output equivalence between implementations

Consider two implementations $(\alpha, \beta, \gamma)$ and $(\hat{\alpha}, \hat{\beta}, \hat{\gamma})$—whether feasible or infeasible—and their respective controllers, $C$ and $\hat{C}$.

By definition, these implementations share identical closed-loop reference signal/control signal transfer functions:

$$T_{ur}^{(\alpha,\beta,\gamma)}(P) = T_{ur}^{(\hat{\alpha},\hat{\beta},\hat{\gamma})}(P) \rightarrow u^{(\alpha,\beta,\gamma)}(P)r = u^{(\hat{\alpha},\hat{\beta},\hat{\gamma})}(P)r \rightarrow u^{(\alpha,\beta,\gamma)}(P) = u^{(\hat{\alpha},\hat{\beta},\hat{\gamma})}(P), \quad (19)$$

and reference/plant output transfer functions:

$$T_{yr}^{(\alpha,\beta,\gamma)}(P) = T_{yr}^{(\hat{\alpha},\hat{\beta},\hat{\gamma})}(P) \rightarrow Pu^{(\alpha,\beta,\gamma)}(P) = Pu^{(\hat{\alpha},\hat{\beta},\hat{\gamma})}(P) \rightarrow u^{(\alpha,\beta,\gamma)}(P) = u^{(\hat{\alpha},\hat{\beta},\hat{\gamma})}(P). \quad (20)$$

This implies that $C$ and $\hat{C}$ are input-output equivalents, or that $C$ and $\hat{C}$ produce *equivalent control actions*, denoted as $C \cong \hat{C}$. Algebraically, the input-output controller equivalence is an equivalence relation, [19, Definition 12.2, p. 303]. This can be summarized in the following property:

**Property 1**: All implementations of a two-parameter stabilizing control law, $u = C_{ff}r - C_{fb}y$—regardless of feasibility—yield input-output equivalence through their respective controllers.

$$\forall P \in \mathcal{R}, \forall (C,\hat{C}) \in \mathcal{S}_{stab}(P) : C \cong \hat{C} \Rightarrow u^{(\alpha,\beta,\gamma)}(P) = u^{(\hat{\alpha},\hat{\beta},\hat{\gamma})}(P)$$

Clearly, this is an intrinsic controller property, independent of the plant $P$. □

For example, the implementations $(C_{ff}, C_{fb})$ (Fig.2) and $(C_r, C_e, C_y)$ (Fig.3), with the YK parametrization $(Q_r, Q_y)$ are input-output equivalent, i.e., $C^{(C_{ff},C_{fb})} \cong C^{(C_r,C_e,C_y)}$.

$$T_{yr}^{(C_{ff},C_{fb})}(P) = \frac{C_{ff}\,P}{1 + C_{fb}\,P} = \frac{\left(\dfrac{Q_r}{X_2 - NQ_y}\right)\left(\dfrac{N}{D}\right)}{1 + \left(\dfrac{X_1 + DQ_y}{X_2 - NQ_y}\right)\left(\dfrac{N}{D}\right)} = \frac{NQ_r}{\underbrace{NX_1 + DX_2}_{1} - DNQ_y + NDQ_y} = NQ_r \quad (21)$$

$$T_{yr}^{(C_r,C_e,C_y)}(P) = \frac{C_r\,C_e\,P}{1 + C_y\,C_e\,P} = \frac{Q_r\left(\dfrac{1}{X_2 - NQ_y}\right)\left(\dfrac{N}{D}\right)}{1 + (X_1 + DQ_y)\left(\dfrac{1}{X_2 - NQ_y}\right)\left(\dfrac{N}{D}\right)} = NQ_r \quad (22)$$

MIMO case

$$T_{yr}^{(C_{ff},C_{fb})}(P) = (I + PC_{fb})^{-1}PC_{ff} = P(I + C_{fb}P)^{-1}C_{ff} \quad (23)$$

$$\begin{aligned}
{T_{yr}^{(C_{ff},C_{fb})}}^{-1} &= \left((I + PC_{fb})^{-1}PC_{ff}\right)^{-1} = \underbrace{C_{ff}^{-1}}_{Q_r^{-1}(X_2 - Q_y\tilde{N})}\ \underbrace{P^{-1}}_{\tilde{N}^{-1}\tilde{D}}\ (I + \underbrace{P}_{\tilde{D}^{-1}\tilde{N}}\ \underbrace{C_{fb}}_{(X_2 - Q_y\tilde{N})^{-1}(X_1 + Q_y\tilde{D})}) \\
&= Q_r^{-1}((X_2 - Q_y\tilde{N})\tilde{N}^{-1}\tilde{D} + (X_1 + Q_y\tilde{D})) = Q_r^{-1}(X_1 + X_2\underbrace{DN^{-1}}_{\tilde{N}^{-1}\tilde{D}}) \\
&= Q_r^{-1}\underbrace{(X_1N + X_2D)}_{I}N^{-1} = Q_r^{-1}N^{-1}
\end{aligned}$$

$$T_{yr}^{(C_{ff},C_{fb})} = NQ_r \quad (24)$$

$$T^{(C_r,C_y,C_e)} = (I + PC_eC_y)^{-1}PC_eC_r = P(I + \underbrace{C_eC_y}_{C_{fb}}P)^{-1}\underbrace{C_eC_r}_{C_{ff}} = P(I + C_{fb}P)^{-1}C_{ff} = NQ_r \quad (25)$$

### G. Input-Output Feedback Implementations

Figure 6 depicts five implementations of the two-parameter stabilizing control law (5) using the input-output feedback topology[12], [13]. The five implementations are input-output equivalent to those previously presented, $(C_{ff}, C_{fb})$ and $(C_r, C_y, C_e)$, through their respective controllers. For both SISO and MIMO systems, these implementations form the core of this work and are analyzed in detail in the following section.

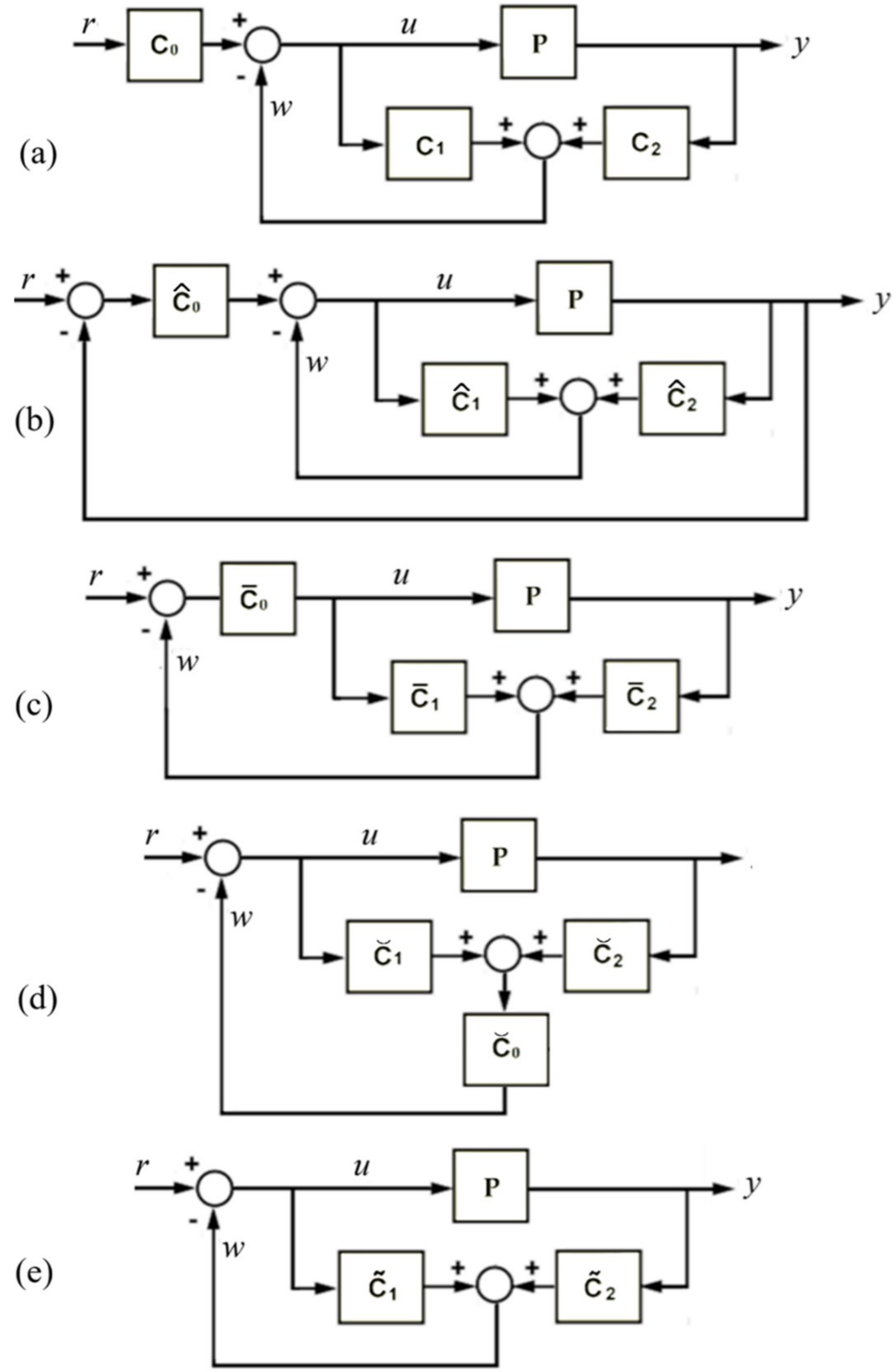

Fig. 6. input-output feedback implementations of the control law (5). (a) input-output feedback with prefilter $(C_0, C_1, C_2)$, (b) two-stage or series cascade input-output feedback $(\hat{C}_0, \hat{C}_1, \hat{C}_2)$, (c) input-output feedback [13] $(\bar{C}_0, \bar{C}_1, \bar{C}_2)$, (d) observer-controller [5] $(\breve{C}_0, \breve{C}_1, \breve{C}_2)$, (e) two-block input-output feedback $(\tilde{C}_1, \tilde{C}_2)$.

The closed-loop transfer functions for the five SISO and MIMO input-output feedback implementations are summarized in Table I.

| SISO case | MIMO case |
|---|---|
| $T_{yr}^{(C_0,C_1,C_2)}(P) = \dfrac{C_0 P}{1 + C_1 + C_2 P}$ | $T_{yr}^{(C_0,C_1,C_2)}(P) = P(I + C_1 + C_2 P)^{-1} C_0$ |
| $T_{yr}^{(\hat{C}_0,\hat{C}_1,\hat{C}_2)}(P) = \dfrac{\hat{C}_0 P}{1 + \hat{C}_1 + \hat{C}_2 P + \hat{C}_0 P}$ | $T_{yr}^{(\hat{C}_0,\hat{C}_1,\hat{C}_2)}(P) = P(I + \hat{C}_1 + \hat{C}_2 P + \hat{C}_0 P)^{-1} \hat{C}_0$ |
| $T_{yr}^{(\bar{C}_0,\bar{C}_1,\bar{C}_2)}(P) = \dfrac{\bar{C}_0 P}{1 + \bar{C}_0 \bar{C}_1 + \bar{C}_0 \bar{C}_2 P}$ | $T_{yr}^{(\bar{C}_0,\bar{C}_1,\bar{C}_2)}(P) = P(I + \bar{C}_0 \bar{C}_1 + \bar{C}_0 \bar{C}_2 P)^{-1} \bar{C}_0$ |
| $T_{yr}^{(\breve{C}_0,\breve{C}_1,\breve{C}_2)}(P) = \dfrac{1}{1 + \breve{C}_0 \breve{C}_1 + \breve{C}_0 \breve{C}_2 P}$ | $T_{yr}^{(\breve{C}_0,\breve{C}_1,\breve{C}_2)}(P) = P(I + \breve{C}_0 \breve{C}_1 + \breve{C}_0 \breve{C}_2 P)^{-1}$ |
| $T_{yr}^{(\tilde{C}_1,\tilde{C}_2)}(P) = P u^{(\tilde{C}_1,\tilde{C}_2)}(P) = \dfrac{P}{1 + \tilde{C}_1 + \tilde{C}_2 P}$ | $T_{yr}^{(\tilde{C}_1,\tilde{C}_2)}(P) = P(I + \tilde{C}_1 + \tilde{C}_2 P)^{-1}$ |

Table I: Closed-loop transfer functions of the five SISO and MIMO I-O feedback implementations.

## III. MAIN RESULTS

In this section, we address the primary objective of this work: demonstrating, within the Youla-Kučera framework, that the constituent blocks of all five I/O feedback implementations in Fig. 6 are proper and stable for both SISO and MIMO systems. For clarity of presentation, we first examine the SISO case. Subsequently, these results are generalized to MIMO plants.

### A. SISO Case

If the plant $P$ does not satisfy the parity interlacing property (PIP), the controller is always unstable. This implies that implementing the two-parameter stabilizing control law (5) with the YK-parameterized two-block controller $C^{(C_{ff},C_{fb})}$ of Fig. 7(a) is infeasible, as both blocks $C_{ff}$ and $C_{fb}$ are unstable , as they share the same unstable poles. On the other hand, implementing this control law is feasible using the standard controller $C^{(C_r,C_e,C_y)}$ of Fig. 7(b), where blocks $C_r$ and $C_y$ are stable, and block $C_e$ is necessarily unstable due to the acyclic topology of the controller.

Note that block $C_e$ can be rewritten as the feedback loop:

$$C_e = \frac{1}{X_2 - NQ_y} = \frac{1}{1 + \underbrace{X_2 - NQ_y - 1}_{C_1}} = \frac{1}{1 + C_1}, \tag{26}$$

where block $C_1$, within the resulting cyclic configuration of Fig. 7(c), is stable.

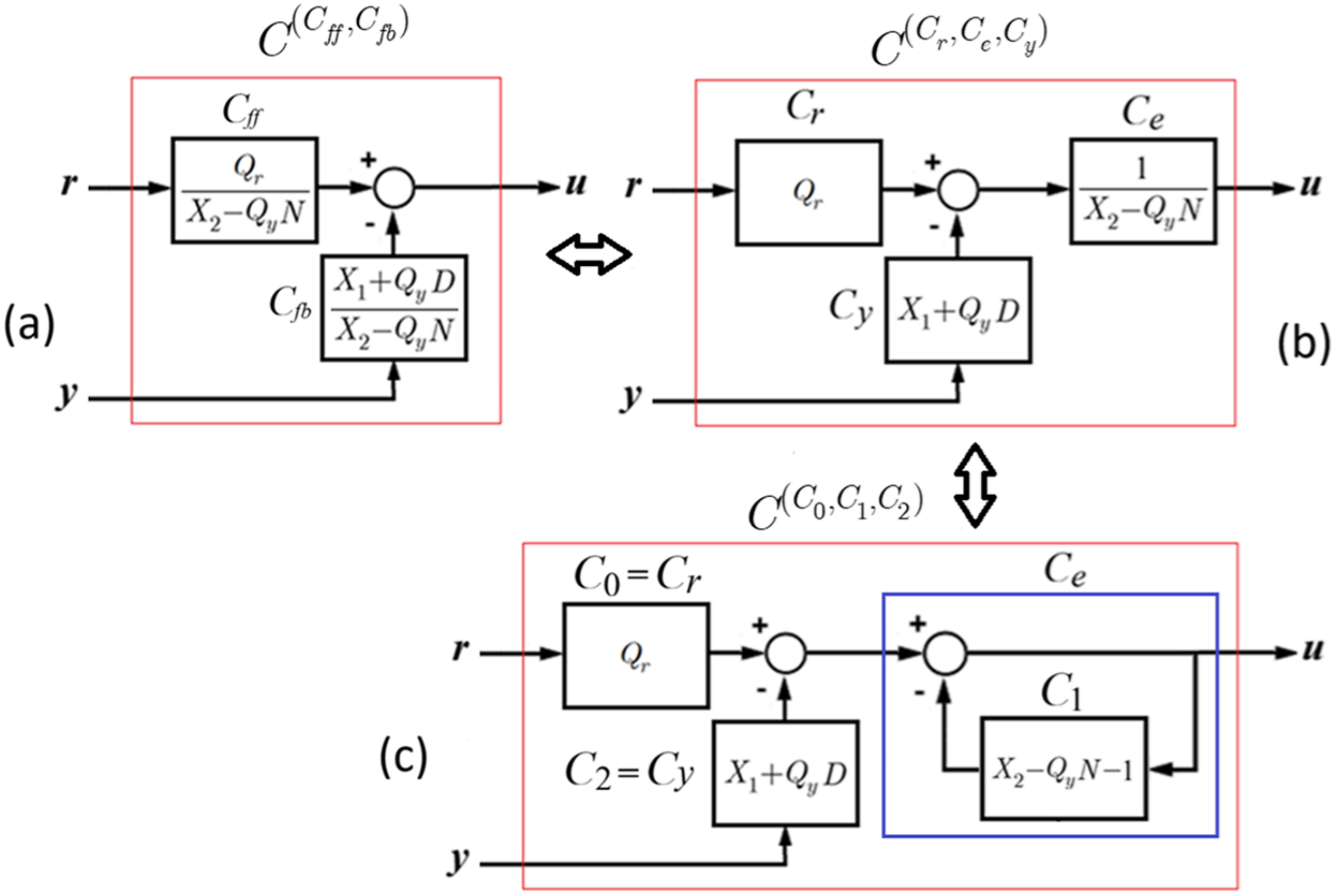

Fig. 7. Three stabilizing controllers (a) Infeasible two-block controller $C^{(C_{ff},C_{fb})}$, (b) Feasible standard controller $C^{(C_r,C_e,C_y)}$, (c) Feasible stable-block controller $C^{(C_0,C_1,C_2)}$.

Consequently, a controller $C^{(C_0,C_1,C_2)}$ with all stable blocks has been obtained; this implementation which we refer to as the input-output feedback with prefilter $(C_0, C_1, C_2)$ (Figs. 6(a) and 8).

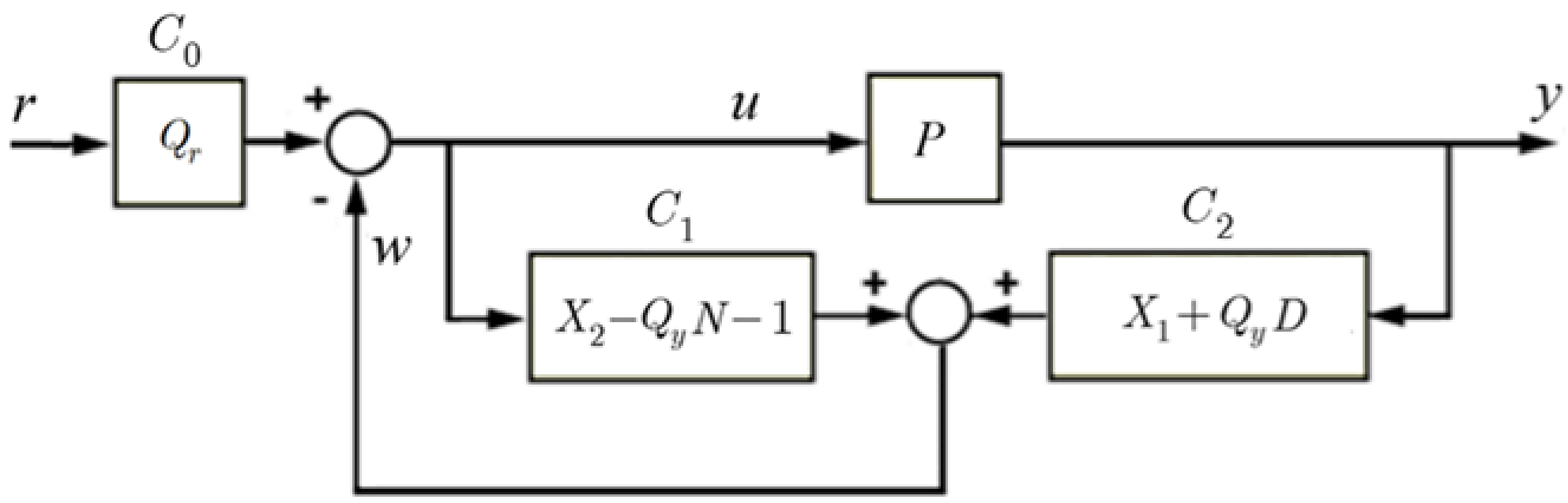

Fig. 8. Input-output feedback with prefilter implementation $(C_0, C_1, C_2)$.

### Input-Output Feedback implementations

Section II-G introduced five implementations based on the input-output feedback topology (Fig. 6). Although the prefiltered structure in Figs. 6(a) and 8 yields proper and stable controller blocks, it is not the only implementation with this property. This section analyzes all five implementations within the input-output feedback family, derives their Youla-Kučera-based parametrizations, and verifies that every topology features proper and stable blocks. Finally, Figure 14 and Table II summarize the parametrizations obtained for the input-output feedback family.

### Input-Output Feedback with Prefilter $(C_0, C_1, C_2)$ implementation

The formal algebraic derivation of YK-based parametrization parametrization of the implementation $(C_0, C_1, C_2)$ is carried-out applying Property 1 to implementations $(C_r, C_e, C_y)$ and $(C_0, C_1, C_2)$.

$$u^{(C_r,C_e,C_y)}(P) = u^{(C_0,C_1,C_2)}(P) \;\Rightarrow\; \frac{C_r C_e}{1 + C_y C_e P} = \frac{C_0}{1 + C_1 + C_2 P}, \tag{27}$$

expanding (27) yields:

$$(C_0 - C_r C_e - C_r C_e C_1) + (C_0 C_y C_e - C_r C_e C_2) P = 0. \tag{28}$$

For equality (28) to hold for any plant $P$, both terms in parentheses must vanish. This results in two equations with three unknowns $C_0$, $C_1$, $C_2$; to ensure a unique solution, the prefilters $C_r$ and $C_0$ from implementations $(C_r, C_y, C_e)$ and $(C_0, C_1, C_2)$ are set equal.

$$\underbrace{C_0 - C_r C_e - C_r C_e C_1}_{0} + \underbrace{C_0 C_y C_e - C_r C_e C_2}_{0} P = 0 \;\underset{C_0 = C_r}{\rightarrow}\; \begin{cases} C_0 = C_r & , \quad C_r = C_0 \\ C_1 = \dfrac{1}{C_e} - 1 & , \quad C_e = \dfrac{1}{C_1 + 1} \\ C_2 = C_y & , \quad C_y = C_2 \end{cases}, \tag{29}$$

from (17) and (29), the conversion between the $(C_r, C_e, C_y)$ and $(C_0, C_1, C_2)$ implementations is given by:

$$\boxed{\begin{array}{lcl} & (C_r, C_e, C_y) \Leftrightarrow (C_0, C_1, C_2) & \\ C_r = C_0 = Q_r & & C_0 = C_r = Q_r \\ C_e = \dfrac{1}{C_1 + 1} = \dfrac{1}{X_2 - N Q_y} & \Leftrightarrow & C_1 = \dfrac{1}{C_e} - 1 = X_2 - N Q_y - 1 \\ C_y = C_2 = X_1 + D Q_y & & C_2 = X_1 + D Q_y \end{array}} \tag{30}$$

Applying Property 1 to implementations $(C_{ff}, C_{fb})$ and $(C_0, C_1, C_2)$.

$$u^{(C_{ff},C_{fb})}(P) = u^{(C_0,C_1,C_2)}(P) \;\Rightarrow\; \frac{C_{ff}}{1 + C_{fb} P} = \frac{C_0}{1 + C_1 + C_2 P}, \tag{31}$$

expanding (31) yields:

$$\underbrace{(C_{ff} + C_{ff}C_1 - C_0)}_{0} + \underbrace{(C_{ff}C_2 - C_0C_{fb})}_{0} P = 0 \rightarrow \begin{cases} C_1 = \dfrac{C_0 - C_{ff}}{C_{ff}} & , \quad C_{ff} = \dfrac{C_0}{C_1 + 1} \\ C_2 = \dfrac{C_0 C_{fb}}{C_{ff}} & , \quad C_{fb} = \dfrac{C_2}{C_1 + 1} \end{cases}, \tag{32}$$

for $C_0 = Q_r$ from (9)

$$C_0 = Q_r = \frac{C_{ff}}{D + NC_{fb}}, \tag{33}$$

substituting $C_{ff}$ and $C_{fb}$ from (9), and (33) into $C_1$ and $C_2$ of (32) yields:

$$(C_{ff}, C_{fb}) \Leftrightarrow (C_0, C_1, C_2)$$

$$\begin{array}{lcl} C_{ff} = \dfrac{C_0}{C_1 + 1} = \dfrac{Q_r}{X_2 - NQ_y} & & C_0 = \dfrac{C_{ff}}{D + NC_{fb}} = Q_r \\ & \Leftrightarrow & C_1 = \dfrac{1 - D - NC_{fb}}{D + NC_{fb}} = X_2 - NQ_y - 1 \\ C_{fb} = \dfrac{C_2}{C_1 + 1} = \dfrac{X_1 + DQ_y}{X_2 - NQ_y} & & C_2 = \dfrac{C_{fb}}{D + NC_{fb}} = X_1 + DQ_y \end{array} \tag{34}$$

For the $(C_0, C_1, C_2)$ implementation in Fig. 9, internal stability is verified by analyzing the responses of $y$, $u$, $y_{C_1}$, $y_{C_2}$ to the exogenous inputs $r$, $d$, and $n$ (the output of the prefilter $C_0$ can be omitted since $Q_r \in \mathcal{S}$).

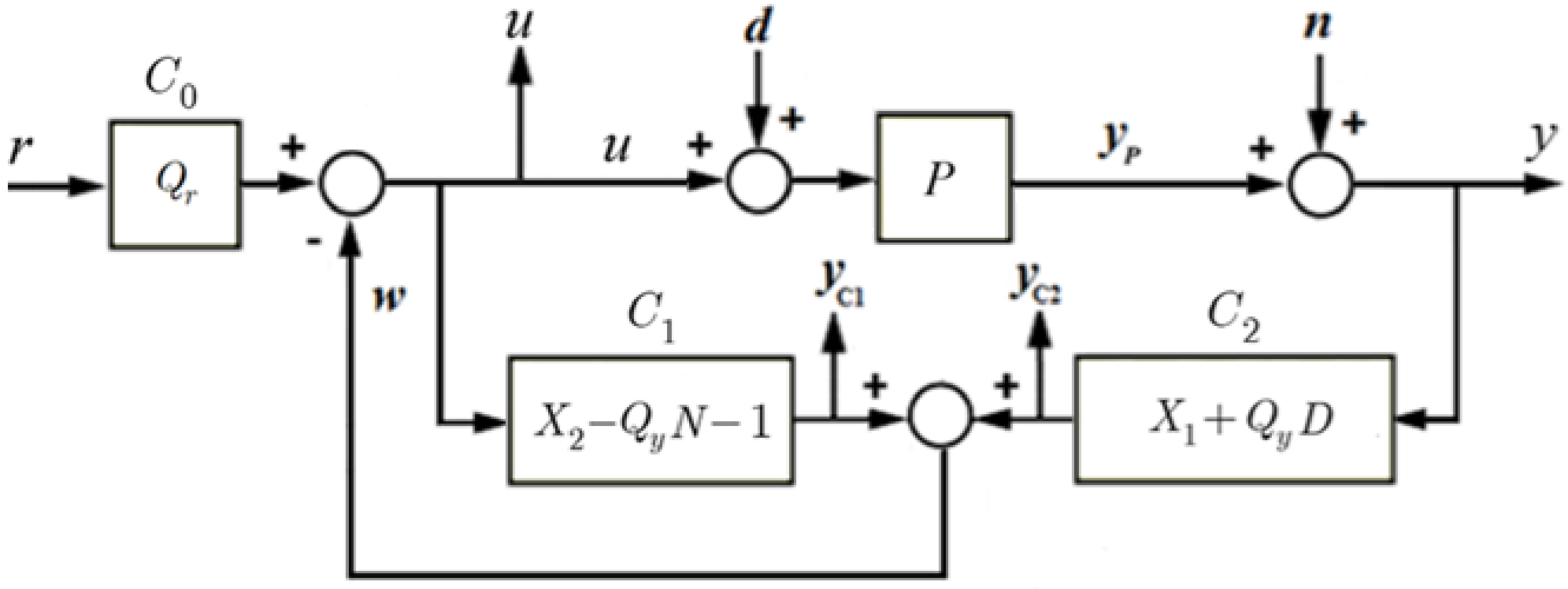

Fig. 9. Input-output feedback with prefilter implementation $(C_0, C_1, C_2)$ with input and output disturbances $d$ and $n$, respectively.

In this implementation, $H^{(C_0,C_1,C_2)}(P,C_0,C_1,C_2)$ is the 4×3 transfer matrix from $[r\ d\ n]^{\mathrm{T}}$ to $[y\ u\ y_{_{C_1}}\ y_{_{C_2}}]^{\mathrm{T}}$.

$$H^{(C_0,C_1,C_2)}(P,C_0,C_1,C_2)=\frac{1}{1+C_1+PC_2}\begin{bmatrix} C_0P & (1+C_1)P & 1+C_1 \\ C_0 & -C_2P & -C_2 \\ C_0C_1 & -C_1C_2P & -C_1C_2 \\ C_0C_2P & C_2(1+C_1)P & C_2(1+C_1) \end{bmatrix}. \tag{35}$$

Taking into account that $C_0=Q_r$, $C_1=X_2-NQ_y-1$, $C_2=X_1+DQ_y$ and the coprime factorization $P=N/D$, $X_1N+X_2D=1$, $N,D,X_1,X_2\in\mathcal{S}$.

$$\Delta^{(C_0,C_1,C_2)}=\frac{1}{1+C_1+PC_2}=\frac{D}{\underbrace{NX_1+DX_2}_{1}}=D, \tag{36}$$

substituting $C_0$. $C_0$, $C_0$ from (34) and (36) into (35), we obtain the matrix $H$ in terms of $N$, $D$ and the free parameters $Q_r$ and $Q_y$.

$$H(N,D,Q_r,Q_y)^{(C_0,C_1,C_2)}=$$
$$\begin{bmatrix} Q_rN & (X_2-Q_yN)N & (X_2-Q_yN)D \\ Q_rD & -(X_1+Q_yD)N & -(X_1+Q_yD)D \\ Q_r(X_2-Q_yN-1)D & -(X_2-Q_yN-1)(X_1+Q_yD)N & -(X_2-Q_yN-1)(X_1+Q_yD)D \\ Q_r(X_1+Q_yD)N & (X_1+Q_yD)(X_2-Q_yN)N & (X_1+Q_yD)(X_2-Q_yN)D \end{bmatrix}. \tag{37}$$

Since $N,D,Q_r,Q_y\in\mathcal{S}$ are stable proper transfer functions, the feedback system is internally stable. Clearly, the transfer functions in the first two rows of (37) (the "gang of six") are identical to those previously obtained in (13). This yields the controller $C^{(C_0,C_1,C_2)}$ parameterized by the free parameters $Q_r$ and $Q_y$, with all its constituent blocks being proper and stable.

**Two-Stage or Series Cascade Input-Output Feedback $(\hat{C}_0,\hat{C}_1,\hat{C}_2)$ implementation**

The blocks $\hat{C}_0$, $\hat{C}_1$ and $\hat{C}_2$ of the two-stage input-output feedback implementation $(\hat{C}_0,\hat{C}_1,\hat{C}_2)$ (Fig. 6(b)) are obtained applying Property 1 to implementations $(C_{ff},C_{fb})$ and $(\hat{C}_0,\hat{C}_1,\hat{C}_2)$.

$$u^{(C_{ff},C_{fb})}(P)=u^{(\hat{C}_0,\hat{C}_1,\hat{C}_2)}(P)\ \Rightarrow\ \frac{C_{ff}}{1+C_{fb}P}=\frac{\hat{C}_0}{1+\hat{C}_1+\hat{C}_2P+\hat{C}_0P}, \tag{38}$$

expanding (38):

$$\underbrace{(C_{ff} + C_{ff}\hat{C}_1 - \hat{C}_0)}_{0} + \underbrace{(C_{ff}\hat{C}_2 + C_{ff}\hat{C}_0 - C_{fb}\hat{C}_0)}_{0} P = 0$$

$$\downarrow$$

$$\begin{cases} \hat{C}_1 = \dfrac{\hat{C}_0}{C_{ff}} - 1,\ \hat{C}_2 = \dfrac{(C_{fb} - C_{ff})\hat{C}_0}{C_{ff}} \\ C_{ff} = \dfrac{\hat{C}_0}{\hat{C}_1 + 1},\ C_{fb} = \dfrac{\hat{C}_2 + \hat{C}_0}{\hat{C}_1 + 1} \end{cases}, \tag{39}$$

substituting $C_{ff}$ and $C_{fb}$ from (9) into $\tilde{C}_1$ and $\tilde{C}_2$ of (39) yields (see Fig. 10):

$$\boxed{\begin{array}{c} (C_{ff}, C_{fb}) \Leftrightarrow (\hat{C}_0, \hat{C}_1, \hat{C}_2) \\ \begin{array}{l} C_{ff} = \dfrac{\hat{C}_0}{\hat{C}_1 + 1} = \dfrac{Q_r}{X_2 - NQ_y} \\ C_{fb} = \dfrac{\hat{C}_2 + \hat{C}_0}{\hat{C}_1 + 1} = \dfrac{X_1 + DQ_y}{X_2 - NQ_y} \end{array} \Leftrightarrow \begin{array}{l} \hat{C}_1 = \dfrac{\hat{C}_0(X_2 - NQ_y) - Q_r}{Q_r} \\ \hat{C}_2 = \dfrac{\hat{C}_0(X_1 + DQ_y)}{Q_r} + \hat{C}_0 \end{array} \Leftrightarrow \begin{array}{l} \hat{C}_0 = Q_r \\ \hat{C}_1 = X_2 - Q_y N - 1 \\ \hat{C}_2 = X_1 + Q_y D - Q_r \end{array} \end{array}} \tag{40}$$

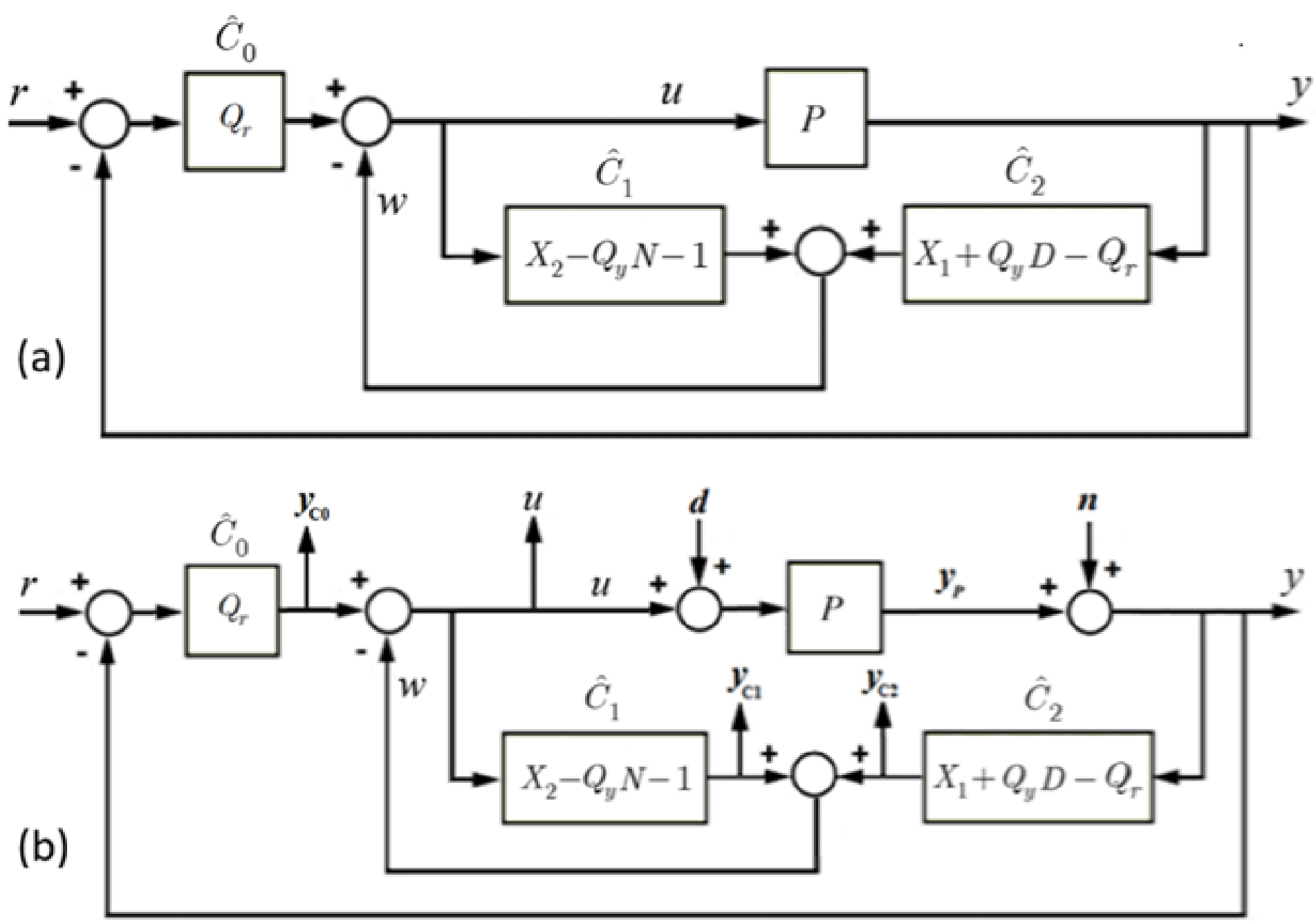

Fig. 10. (a) Two-Stage or Series Cascade Input-Output Feedback implementation $(\hat{C}_0, \hat{C}_1, \hat{C}_2)$,
(b) Implementation with input and output disturbances $d$ and $n$, respectively.

The matrix $H$ for output signals $y$, $u$, $y_{C_0}$, $y_{C_1}$, $y_{C_2}$ under inputs signals $r$, $d$ and $n$ is:

$$H^{(\hat{C}_0,\hat{C}_1,\hat{C}_2)}(N,D,Q_r,Q_y) = \begin{bmatrix} Q_r N & (X_2 - Q_y N)N & (X_2 - Q_y N)D \\ Q_r D & -(X_1 + Q_y D)N & -(X_1 + Q_y D)D \\ Q_r(1 - Q_r N) & -Q_r(X_2 - Q_y N)N & -Q_r(X_2 - Q_y N)D \\ Q_r(X_2 - Q_y N - 1)D & -(X_2 - Q_y N - 1)(X_1 + Q_y D)N & -(X_2 - Q_y N - 1)(X_1 + Q_y D)D \\ Q_r(X_1 + Q_y D - Q_r)N & (X_1 + Q_y D - Q_r)(X_2 - Q_y N)N & (X_1 + Q_y D - Q_r)(X_2 - Q_y N)D \end{bmatrix} \quad (41)$$

**Input-Output Feedback $(\bar{C}_0,\bar{C}_1,\bar{C}_2)$ Implementation**

Consider the implementation of two-parameter stabilizing control law (5) with the Input-Output Feedback implementation $(\bar{C}_0,\bar{C}_1,\bar{C}_2)$ of Fig. 6(c). From input-output equivalence (Property 1) between $(\bar{C}_0,\bar{C}_1,\bar{C}_2)$ and $(C_{ff},C_{fb})$.

$$u^{(\bar{C}_0,\bar{C}_1,\bar{C}_2)}(P) = u^{(C_{ff},C_{fb})}(P) \rightarrow \frac{\bar{C}_0}{1 + \bar{C}_0\bar{C}_1 + \bar{C}_0\bar{C}_2 P} = \frac{C_{ff}}{1 + C_{fb}P}, \quad (42)$$

rearranging (42) yields:

$$\underbrace{(C_{ff} + C_{ff}\bar{C}_0\bar{C}_1 - \bar{C}_0)}_{0} + \bar{C}_0\underbrace{(C_{ff}\bar{C}_2 - C_{fb})}_{0}P = 0, \quad (43)$$

substituting $\bar{C}_0 = C_0$ and the YK parametrization of $C_{ff}$ and $C_{fb}$ (9) into (43) yields (see Fig. 11(a)):

$$(C_{ff},C_{fb}) \Leftrightarrow (\bar{C}_0,\bar{C}_1,\bar{C}_2)$$

$$\begin{aligned} C_{ff} &= \frac{\bar{C}_0}{1+\bar{C}_1\bar{C}_0} = \frac{Q_r}{X_2 - Q_y N} \\ C_{fb} &= \frac{\bar{C}_0\bar{C}_2}{1+\bar{C}_1\bar{C}_0} = \frac{X_1 + Q_y D}{X_2 - Q_y N} \end{aligned} \Leftrightarrow \begin{aligned} \bar{C}_1 &= \frac{C_0 - C_{ff}}{C_0 C_{ff}} = \frac{X_2 - Q_y N - 1}{Q_r} \\ \bar{C}_2 &= \frac{C_{fb}}{C_{ff}} = \frac{X_1 + Q_y D}{Q_r} \end{aligned} \Leftrightarrow \begin{aligned} \bar{C}_0 &= Q_r \in \mathcal{U} \\ \bar{C}_1 &= \frac{X_2 - Q_y N - 1}{Q_r} \\ \bar{C}_2 &= \frac{X_1 + Q_y D}{Q_r} \end{aligned} \quad (44)$$

Blocks $\bar{C}_1$ and $\bar{C}_2$ will be stable only if the free parameter $Q_r$ is bistable ($Q_r \in \mathcal{U}$ or $Q_r, Q_r^{-1} \in \mathcal{S}$). This constraint is not severe from a practical standpoint.

Regarding the implementation $(\bar{C}_0,\bar{C}_1,\bar{C}_2)$ in Fig. 11(b), internal stability is verified as before through the analysis of responses of $y$, $u$, $y_p$, $y_{C_1}$, and $y_{C_2}$ to the exogenous inputs $r$, $d$ and $n$.

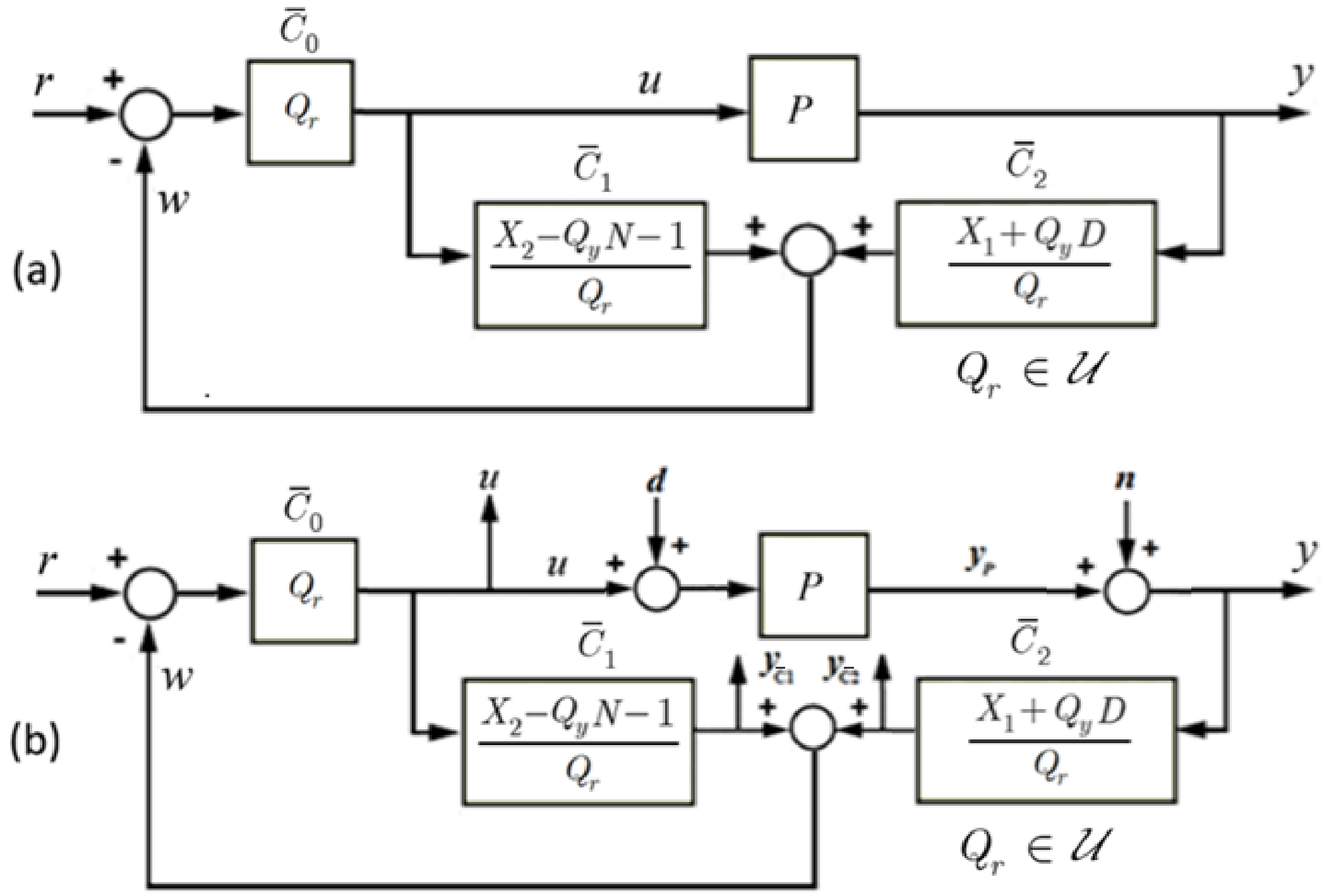

Fig. 11. (a) Input-output feedback implementation $(\bar{C}_0,\bar{C}_1,\bar{C}_2)$, (b) Implementation with input and output disturbances $d$ and $n$, respectively.

By repeating the previous calculation, the matrix $H$ for the $(\bar{C}_0,\bar{C}_1,\bar{C}_2)$ implementation is obtained in terms of the plant $P = N / D$ and the stable free parameters $Q_r \in \mathcal{U}, Q_y$.

$$H^{(\bar{C}_0,\bar{C}_1,\bar{C}_2)}(N,D,Q_r,Q_y) =$$

$$\begin{bmatrix} Q_r N & (X_2 - Q_y N)N & (X_2 - Q_y N)D \\ Q_r D & -(X_1 + Q_y D)N & -(X_1 + Q_y D)D \\ (X_2 - Q_y N - 1)D & \dfrac{-(X_1 + Q_y D)(X_2 - Q_y N - 1)N}{Q_r} & -\dfrac{(X_1 + Q_y D)(X_2 - Q_y N - 1)D}{Q_r} \\ (X_1 + Q_y D)N & \dfrac{(X_1 + Q_y D)(X_2 - Q_y N)N}{Q_r} & \dfrac{(X_1 + Q_y D)(X_2 - Q_y N)D}{Q_r} \end{bmatrix}. \quad (45)$$

The properness and stability of $N, D, Q_r, Q_r^{-1}.Q_y$ ensures that all entries of $H(N,D,Q_r,Q_y)$ are proper and stable.

**Two-Block $(\tilde{C}_1,\tilde{C}_2)$ and Observer-Controller $(\breve{C}_0,\breve{C}_1,\breve{C}_2)$ Input-Output Feedback implementations**

The blocks $\tilde{C}_1$ y $\tilde{C}_2$ of the two-block input-output feedback implementation $(\tilde{C}_1,\tilde{C}_2)$ (Fig. 6(e)) are obtained applying Property 1 to implementations $(C_{ff},C_{fb})$ and $(\tilde{C}_1,\tilde{C}_2)$.

$$u^{(C_{ff},C_{fb})}(P) = u^{(\tilde{C}_1,\tilde{C}_2)}(P) \;\Rightarrow\; \frac{C_{ff}}{1+C_{fb}P} = \frac{1}{1+\tilde{C}_1+\tilde{C}_2P}, \tag{46}$$

expanding (46): $$\underbrace{C_{ff}(1+\tilde{C}_1)-1}_{0} + \underbrace{(C_{ff}\tilde{C}_2 - C_{fb})}_{0}P = 0 \rightarrow \begin{cases} \tilde{C}_1 = \dfrac{1}{C_{ff}} - 1,\; \tilde{C}_2 = \dfrac{C_{fb}}{C_{ff}} \\ C_{ff} = \dfrac{1}{\tilde{C}_1+1},\; C_{fb} = \dfrac{\tilde{C}_2}{\tilde{C}_1+1} \end{cases}, \tag{47}$$

substituting $C_{ff}$ and $C_{fb}$ from (9) into $\tilde{C}_1$ and $\tilde{C}_2$ of (47) yields:

$$\boxed{\begin{array}{c} (C_{ff},C_{fb}) \Leftrightarrow (\tilde{C}_1,\tilde{C}_2) \\ \begin{array}{l} C_{ff} = \dfrac{1}{\tilde{C}_1+1} = \dfrac{Q_r}{X_2-NQ_y} \\ C_{fb} = \dfrac{\tilde{C}_2}{\tilde{C}_1+1} = \dfrac{X_1+DQ_y}{X_2-NQ_y} \end{array} \Leftrightarrow \begin{array}{l} \tilde{C}_1 = \dfrac{1}{C_{ff}} - 1 = \dfrac{X_2-NQ_y-Q_r}{Q_r} \\ \tilde{C}_2 = \dfrac{C_{fb}}{C_{ff}} = \dfrac{X_1+DQ_y}{Q_r} \end{array}, \; Q_r \in \mathcal{U} \end{array}} \tag{48}$$

Figure 12 illustrates the relationship between the Two-Block $(\tilde{C}_1,\tilde{C}_2)$ and the Observer-Controller $(\breve{C}_0,\breve{C}_1,\breve{C}_2)$ input-output feedback implementations.

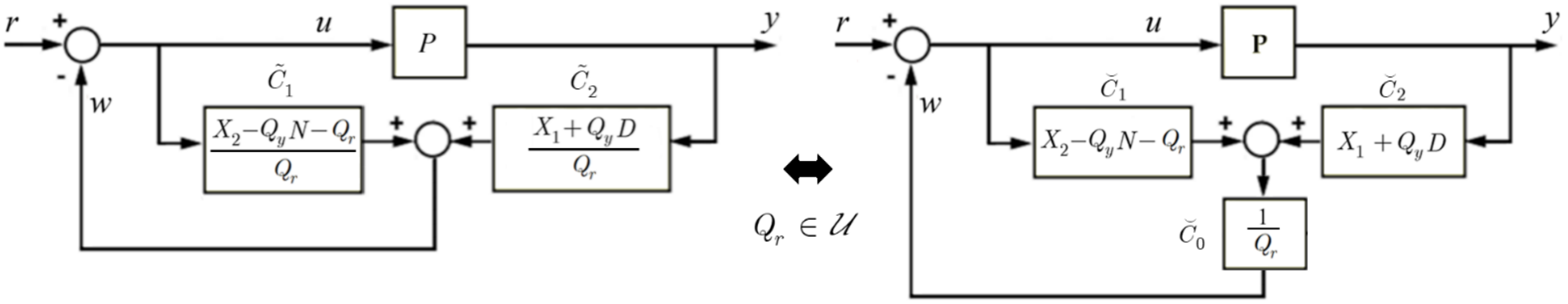

Fig. 12. (a) Two-Block $(\tilde{C}_1,\tilde{C}_2)$ input-output feedback implementation, (b) Observer-Controller implementation $(\breve{C}_0,\breve{C}_1,\breve{C}_2)$.

$$\boxed{\begin{array}{c} (C_{ff},C_{fb}) \Leftrightarrow (\breve{C}_0,\breve{C}_1,\breve{C}_2) \\ \begin{array}{l} C_{ff} = \dfrac{1}{\breve{C}_0\breve{C}_1+1} = \dfrac{Q_r}{X_2-NQ_y} \\ C_{fb} = \dfrac{\breve{C}_0\breve{C}_2}{\breve{C}_0\breve{C}_1+1} = \dfrac{X_1+DQ_y}{X_2-NQ_y} \end{array} \Leftrightarrow \begin{array}{l} \breve{C}_1 = \dfrac{1-C_{ff}}{\breve{C}_0C_{ff}} = \dfrac{X_2-NQ_y-Q_r}{\breve{C}_0Q_r} \\ \breve{C}_2 = \dfrac{C_{fb}}{\breve{C}_0C_{ff}} = \dfrac{X_1+DQ_y}{\breve{C}_0Q_r} \end{array} \Leftrightarrow \begin{array}{l} \breve{C}_0 = \dfrac{1}{Q_r} \\ \breve{C}_1 = X_2-NQ_y-Q_r \\ \breve{C}_2 = X_1+DQ_y \end{array} \\ Q_r \in \mathcal{U} \end{array}} \tag{49}$$

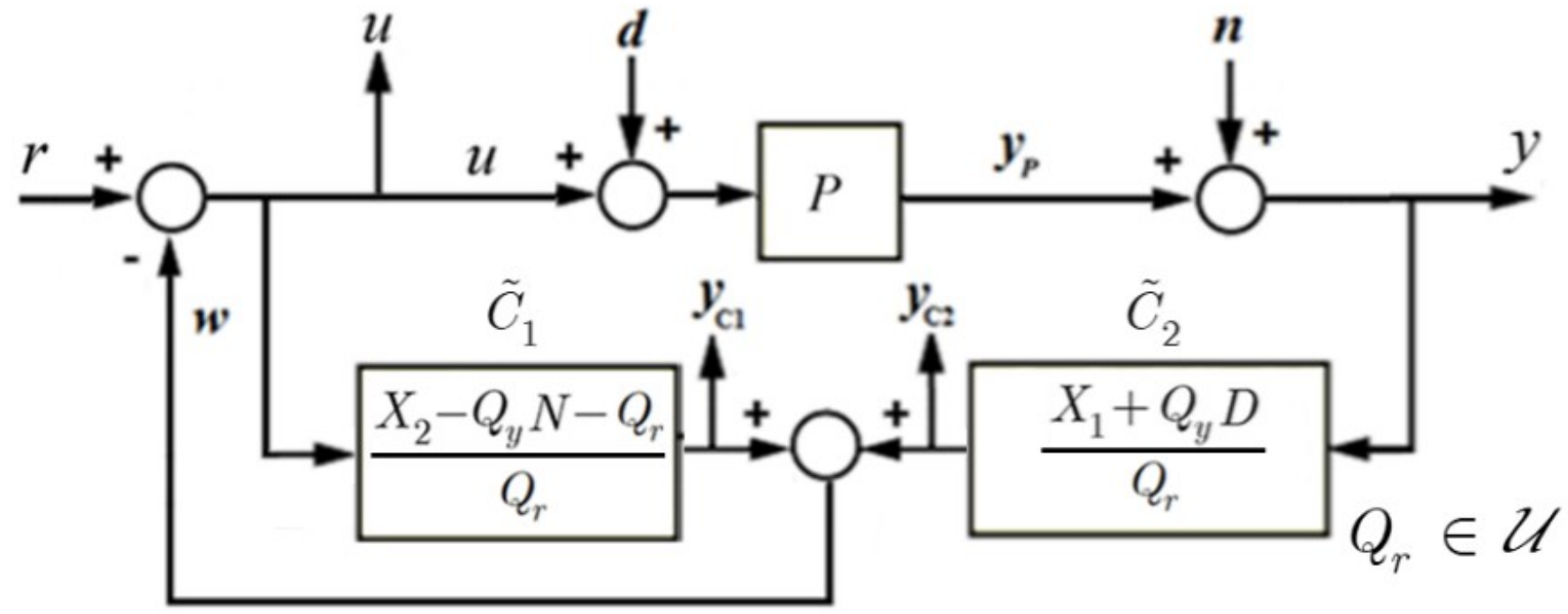

Fig. 13. Two-Block $(\tilde{C}_1, \tilde{C}_2)$ input-output feedback implementation with input and output disturbances $d$ and $n$, respectively.

From Fig. 13, the matrix $H$ for output signals $y, u, y_{c_0}, y_{c_1}$, and $y_{c_2}$ to the exogenous inputs $r$, $d$ and $n$ is:

$$H(N, D, Q_r, Q_y) = \begin{bmatrix} Q_r N & (X_2 - Q_y N)N & (X_2 - Q_y N)D \\ Q_r D & -(X_1 + Q_y D)N & -(X_1 + Q_y D)D \\ (X_2 - Q_y N - Q_r)D & \dfrac{-(X_2 - Q_y N - Q_r)(X_1 + Q_y D)N}{Q_r} & \dfrac{-(X_2 - Q_y N - Q_r)(X_1 + Q_y D)N}{Q_r} \\ (X_1 + Q_y D)N & \dfrac{(X_2 - Q_y N)(X_1 + Q_y D)N}{Q_r} & \dfrac{(X_2 - Q_y N)(X_1 + Q_y D)D}{Q_r} \end{bmatrix}. \quad (50)$$

This implementation is stabilizing for $Q_r \in \mathcal{U}$ and arbitrary $Q_y \in \mathcal{S}$.

For an arbitrary plant $P \in \mathcal{R}$, Kwok and Davison [18] correctly conjecture that no two-block implementation of the control law $u = C_{ff} r - C_{fb} y$ is universally feasible. The authors computationally evaluated $3^{12} = 531,441$ distinct two-block configurations for $Q_r, Q_y \in \mathcal{S}$. Conversely, the two-block implementation $(\tilde{C}_1, \tilde{C}_2)$ is universally feasible, albeit restricted to $Q_r \in \mathcal{U} \subset \mathcal{S}$ and $Q_y \in \mathcal{S}$.

Figure 14 and Table II summarize the family of five SISO input-output feedback implementations for the control law $u = C_{ff} r - C_{fb} y$. As can be observed, the constituent blocks of all implementations are stable for $Q_r \in \mathcal{U}$.

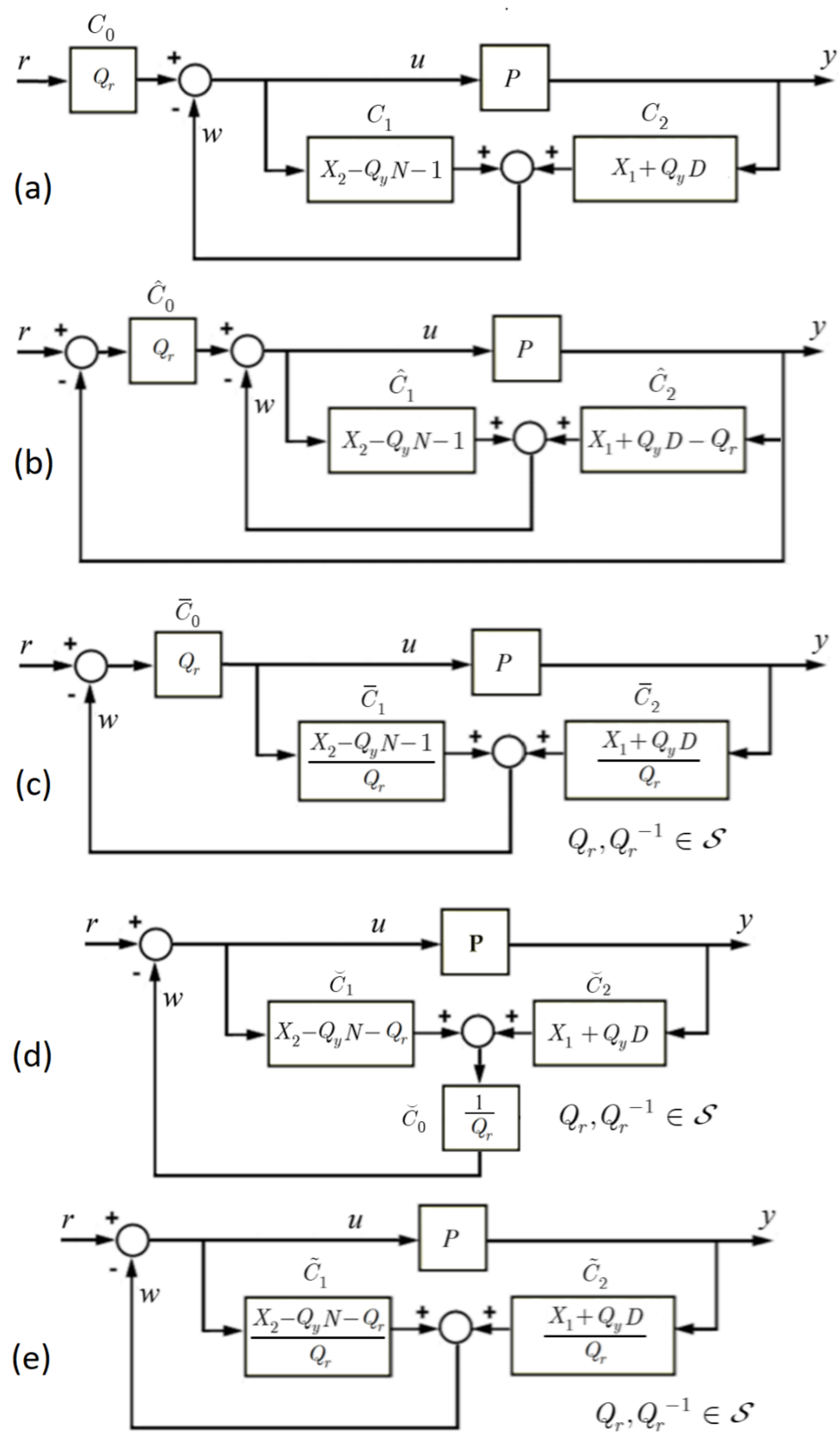

Fig. 14. Stabilization of SISO systems with Input-output feedback implementations. (a) input-output feedback with prefilter $(C_0, C_1, C_2)$, (b) two-stage input-output feedback $(\hat{C}_0, \hat{C}_1, \hat{C}_2)$, (c) input-output feedback $(\bar{C}_0, \bar{C}_1, \bar{C}_2)$, (d) observer-controller $(\breve{C}_0, \breve{C}_1, \breve{C}_2)$, (e) two-block input-output feedback $(\tilde{C}_1, \tilde{C}_2)$.

| Implementation | Blocks | | |
|---|---|---|---|
| $(C_0, C_1, C_2)$ | $C_0 = Q_r$ | $C_1 = X_2 - NQ_y - 1$ | $C_2 = X_1 + DQ_y$ |
| $(\hat{C}_0, \hat{C}_1, \hat{C}_2)$ | $\hat{C}_0 = Q_r$ | $\hat{C}_1 = X_2 - Q_y N - 1$ | $\hat{C}_2 = X_1 + Q_y D - Q_r$ |
| $(\bar{C}_0, \bar{C}_1, \bar{C}_2)$ | $\bar{C}_0 = Q_r$ | $\bar{C}_1 = \dfrac{X_2 - Q_y N - 1}{Q_r}$ | $\bar{C}_2 = \dfrac{X_1 + Q_y D}{Q_r}$ |
| $(\breve{C}_0, \breve{C}_1, \breve{C}_2)$ | $\breve{C}_0 = \dfrac{1}{Q_r}$ | $\breve{C}_1 = X_2 - NQ_y - Q_r$ | $\breve{C}_2 = X_1 + DQ_y$ |
| $(\tilde{C}_1, \tilde{C}_2)$ | ——— | $\tilde{C}_1 = \dfrac{X_2 - NQ_y - Q_r}{Q_r}$ | $\tilde{C}_2 = \dfrac{X_1 + DQ_y}{Q_r}$ |

Table II: Parametrization of the five SISO input-output feedback implementations

The restriction $Q_r \in \mathcal{U}$ (or $Q_r, Q_r^{-1} \in \mathcal{S}$) in implementations $(\bar{C}_0, \bar{C}_1, \bar{C}_2)$, $(\breve{C}_0, \breve{C}_1, \breve{C}_2)$, and $(\tilde{C}_0, \tilde{C}_1)$ is not severe from a practical standpoint. In what follows, the conditions under which $Q_r \in \mathcal{U}$ are analyzed.

**Analysis of $Q_r$ bistable ($Q_r \in \mathcal{U} \subset \mathcal{S}$ or $Q_r, Q_r^{-1} \in \mathcal{S}$)**

The parameter $Q_r$ does not influence system stability through $C_{fb}$; it only affects the reference input response through $C_{ff}$.

Considering that:

$$T_{yr} = \frac{n_{T_{yr}}}{d_{T_{yr}}}, P = \frac{N}{D} = \frac{\frac{n_P}{d}}{\frac{d_P}{d}} = \frac{n_P}{d_P}, N = \frac{n_P}{d}, D = \frac{d_P}{d}, \deg(d) = \deg(d_P), \rho(N) = \rho(P), d \in H. \quad (51)$$

All implementations of the two-parameter control law (5) have the same closed-loop transfer function (13) from input $r$ to output $y$.

$$T_{yr} = Q_r N. \quad (52)$$

Where $N$ represents the physical limitations imposed by the RHP zeros of the plant, and $Q_r$ represents the remaining design freedom. If $Q_r$ is bistable, the following conditions must be met:

i) From (52), $T_{yr}$ must contain all the RHP zeros of $P$ (which are the same as those of $N$) and must not add additional RHP zeros to those already present in $N$ to prevent $Q_r$ from having unstable poles, i.e., $Q_r \in \mathcal{S}$.

$$n_P = n_P^{RHP} n_P^{LHP},\ n_{T_{yr}} = n_P^{RHP} n_{T_{yr}}^{LHP}$$

$$Q_r = \frac{T_{yr}}{N} = \frac{n_{T_{yr}}}{n_P}\frac{d_N}{d_{T_{yr}}} = \frac{n_{T_{yr}}^{LHP}}{n_P^{LHP}}\frac{d_N}{d_{T_{yr}}}$$

$$Q_r^{-1} = \frac{N}{T_{yr}} = \frac{n_P^{LHP}}{n_{T_{yr}}^{LHP}}\frac{d_{T_{yr}}}{d_N}$$

ii) The relative degree of $T_{yr}$ must be equal to that of the plant $P$ so that $\rho(Q_r) = 0$ (since $Q_r$ is biproper).

$$\rho(T_{yr}) = \underbrace{\rho(Q_r)}_{0} + \underbrace{\rho(N)}_{\rho(P)} = \rho(P).$$

From a design perspective, condition (ii) is related to the plant's minimum-phase zeros $z_{LHP}^G$.

If $\deg(d_{T_{yr}}) = \deg(d) = \deg(d_P)$, then according to condition (i), the numerator of $T_{yr}$ must contain all RHP zeros, and according to condition (ii), it must have the same number of LHP zeros as the plant (if it has fewer or more minimum-phase zeros than the plant, $\rho(T_{yr}) \neq \rho(P)$). This is typically the case when no LHP zeros of the plant are canceled. If some of the plant's LHP zeros are canceled, either the same number of minimum-phase zeros must be added to $T_{yr}$ at suitable locations, or the order of $T_{yr}$ must be reduced $(\deg(d_{T_{yr}}) = \deg(d_P) - \#\,\text{canceled}\,LHP\,\text{zeros}))$ in order to maintain $\rho(T_{yr}) = \rho(P)$ (see Example 2).

If, for some reason, the designer chooses a free parameter $Q_r$ that is not bistable—for instance, if they wish to cancel a minimum-phase zero of the plant without replacing it and without reducing the order of $T_{yr}$—the three-block implementations $(C_0, C_1, C_2)$ or $(\hat{C}_0, \hat{C}_1, \hat{C}_2)$ can be used instead of $(\bar{C}_0, \bar{C}_1, \bar{C}_2)$. However, there is no alternative implementation to $(\tilde{C}_1, \tilde{C}_2)$ using only two blocks.

**Application Examples for SISO Systems**

Since the primary focus of this work is the implementation of two-parameter stabilizing controllers using stable blocks rather than a novel synthesis method, the following examples are intended solely to validate the proposed framework.

**Example 1**

a) Consider the following continuous-time plant that does not satisfy the parity interlacing property (PIP); therefore, it does not admit a stable stabilizing controller.

$$P(s) = \frac{(s-1)}{s(s-2)}$$

The adopted coprime factorization of $P(s)$ is:

$$N(s) = \frac{(s-1)}{(s+2)^2},\ D(s) = \frac{s(s-2)}{(s+2)^2},\ X_1(s) = \frac{36s-8}{s+2},\ X_2(s) = \frac{s-28}{s+2}.$$

A model matching design is carried out for the desired input-output closed loop model:

$$P_m(s) = -\frac{(s-1)}{(s+1)^2}$$

Following the procedure described by Chen [25, p. 405], $C_{ff}(s)$, $C_{fb}(s)$ are obtained, while the free parameters $Q_r(s)$, $Q_y(s)$ follow from (9).

$$\begin{aligned} C_{ff}(s) &= \frac{Q_r(s)}{X_2(s) - Q_y(s)N(s)} = \frac{-s-10}{s-45} \\ C_{fb}(s) &= \frac{X_1(s) + Q_y(s)D(s)}{X_2(s) - Q_y(s)N(s)} = \frac{59s-10}{s-45} \end{aligned} \quad \Leftrightarrow \quad \begin{aligned} Q_r(s) &= \frac{C_{ff}(s)}{D(s) + N(s)C_{fb}(s)} = -1 \\ Q_y(s) &= \frac{X_2(s)C_{fb}(s) - X_1(s)}{D(s) + N(s)C_{fb}(s)} = \frac{45(s+1)}{(s+10)} \end{aligned}.$$

Note that since no LHP zero of the plant $P(s)$ is canceled, $Q_r(s)$ is bistable.

The following table summarize the parametrization of the five SISO input-output feedback implementations

| Implementation | Blocks | | |
|---|---|---|---|
| $(C_0, C_1, C_2)$ | $C_0(s) = -1$ | $C_1(s) = -\frac{55}{s+10}$ | $C_2(s) = \frac{59s-10}{s+10}$ |
| $(\hat{C}_0, \hat{C}_1, \hat{C}_2)$ | $\hat{C}_0(s) = -1$ | $\hat{C}_1(s) = \frac{-55}{s+10}$ | $\hat{C}_2(s) = \frac{60s}{s+10}$ |
| $(\bar{C}_0, \bar{C}_1, \bar{C}_2)$ | $\bar{C}_0(s) = -1$ | $\bar{C}_1(s) = \frac{55}{s+10}$ | $\bar{C}_2(s) = \frac{-59s+10}{s+10}$ |
| $(\breve{C}_0, \breve{C}_1, \breve{C}_2)$ | $\breve{C}_0(s) = -1$ | $\breve{C}_1(s) = \frac{2s-35}{s+10}$ | $\breve{C}_2(s) = \frac{59s-10}{s+10}$ |
| $(\tilde{C}_1, \tilde{C}_2)$ | ——— | $\tilde{C}_1(s) = \frac{-2s+35}{s+10}$ | $\tilde{C}_2(s) = \frac{-59s+10}{s+10}$ |

All implementations have the same "gang of six" transfer functions (13):

$$\begin{cases} T_{yr}(s) = -\dfrac{(s-1)}{(s+1)^2}, \; T_{yd}(s) = \dfrac{(s-1)(s-45)}{(s+1)^2(s+10)}, \qquad T_{yn}(s) = \dfrac{s(s-2)(s-45)}{(s+1)^2(s+10)} \\ T_{ur}(s) = -\dfrac{s(s-2)}{(s+1)^2}, \; T_{ud}(s) = -\dfrac{59(s-0.17)(s-1)}{(s+1)^2(s+10)}, \; T_{un}(s) = -\dfrac{59\,s\,(s-0.17)(s-2)}{(s+1)^2(s+10)} \end{cases}$$

b) In this part of the example, the plant $P(s)$ and the controller $C^{(C_0,C_1,C_2)}$ are discretized. The discretization sampling period $T_s$ is selected to ensure a sufficiently high sampling rate, allowing the digital controller to react before the unstable dynamics grow unbounded, and before the phase distortion introduced by the non-minimum phase (NMP) zero renders the discrete-time system completely unstabilizable. To balance these competing effects, the critical frequency of the right-half plane is computed as:

$$\omega_{max} = \max\left(|\, p_i \,|, |\, z_i \,|\right) \text{ for all } \mathrm{Re}(p_i),\, \mathrm{Re}(z_i) > 0 \to \omega_{max} = \max(2,1) = 2\,\mathrm{rad/s}$$

The sampling frequency $\omega_s$ is selected to be between 10 and 20 times the critical frequency $\omega_{max}$.

$$\omega_s \in \left[10\omega_{max}, 20\omega_{max}\right] = \left[20\,\mathrm{rad/s}, 40\,\mathrm{rad/s}\right]$$

The discretization sampling period is chosen to lie within the interval:

$$T_s = \frac{2\pi}{\omega_s} \in \left[\frac{\pi}{10\omega_s}, \frac{\pi}{5\omega_s}\right]\left[\frac{0.31}{\omega_s}, \frac{0.63}{\omega_s}\right] = \left[0.16\,\mathrm{s},\, 0.31\,\mathrm{s}\right]$$

The intermediate value of $T_s = 0.2\,\mathrm{s}$ is selected.

The $z$-domain transfer functions are:

$$\hat{P}(z) = \frac{0.223\,(z-1.22)}{(z-1)\,(z-1.49)}, \; \hat{P}_m(z) = -\frac{0.15\,(z-1.23)}{(z-0.82)^2}$$

$$\hat{N}(z) = \frac{0.15\,(z-1.23)}{(z-0.82)^2}, \; \hat{D}(z) = \frac{(z-1)(z-1.31)}{(z-0.82)^2}, \; \hat{X}_1(s) = \frac{14\,(z-1.013)}{(z-0.82)}, \; \hat{X}_2(z) = \frac{(z-2.63)}{(z-0.82)}$$

$$\hat{Q}_r(z) = -1, \; \hat{Q}_y(z) = \frac{45\,(z-0.91)}{(z-0.14)}$$

Parametrization of the five discrete-time SISO input-output feedback implementations

| Implementation | Blocks | | |
|---|---|---|---|
| $(C_0^d, C_1^d, C_2^d)$ | $C_0^d(z) = -1$ | $C_1^d(z) = -\dfrac{4.76}{z-0.14}$ | $C_2^d(z) = \dfrac{59\ (z-1.06)}{(z-0.14)}$ |
| $(\hat{C}_0^d, \hat{C}_1^d, \hat{C}_2^d)$ | $\hat{C}_0^d(z) = -1$ | $\hat{C}_1^d(z) = -\dfrac{4.76}{z-0.14}$ | $\hat{C}_2^d(z) = \dfrac{60\,(z-1)}{(z-0.14)}$ |
| $(\bar{C}_0, \bar{C}_1, \bar{C}_2)$ | $\bar{C}_0^d(z) = -1$ | $\bar{C}_1^d(z) = \dfrac{4.76}{z-0.14}$ | $\bar{C}_2^d(z) = -\dfrac{59\,(z-1.015)}{(z-0.14)}$ |
| $(\breve{C}_0, \breve{C}_1, \breve{C}_2)$ | $\breve{C}_0^d(z) = -1$ | $\breve{C}_1^d(z) = \dfrac{2\ (z-2.51)}{(z-0.14)}$ | $\breve{C}_2^d(z) = \dfrac{59\,(z-1.015)}{(z-0.14)}$ |
| $(\tilde{C}_1, \tilde{C}_2)$ | ———— | $\tilde{C}_1^d(z) = -\dfrac{2\ (z-2.51)}{(z-0.14)}$ | $\tilde{C}_2^d(z) = -\dfrac{59\ (z-1.015)}{(z-0.14)}$ |

The "gang of six" discrete transfer functions are:

$$\begin{cases} \hat{T}_{yr}(z) = \dfrac{-0.15\,(z-1.23)}{(z-0.82)^2}, \quad \hat{T}_{yd}(z) = \dfrac{-0.35\,(z+0.66)(z-1.22)}{(z-0.82)^2(z-0.14)}, \quad \hat{T}_{yn}(z) = \dfrac{(z-1)(z^2-4z+4.33)}{(z-0.82)^2(z-0.14)} \\ \hat{T}_{ur}(z) = \dfrac{-(z-1)(z-1.31)}{(z-0.82)^2}, \hat{T}_{ud}(z) = \dfrac{-3.24\,(z-1.267)(z-1.033)}{(z-0.82)^2(z-0.14)}, \hat{T}_{un}(z) = \dfrac{-59\,(z-1)(z^2-1.92z+0.95)}{(z-0.82)^2(z-0.14)} \end{cases}$$

The following figure shows, for the $(C_0, C_1, C_2)$ implementation, the step response from the reference input $r$ to the plant output $y$ (top), and from $r$ to the control input $u$ (bottom), for both the continuous-time and discrete-time cases.

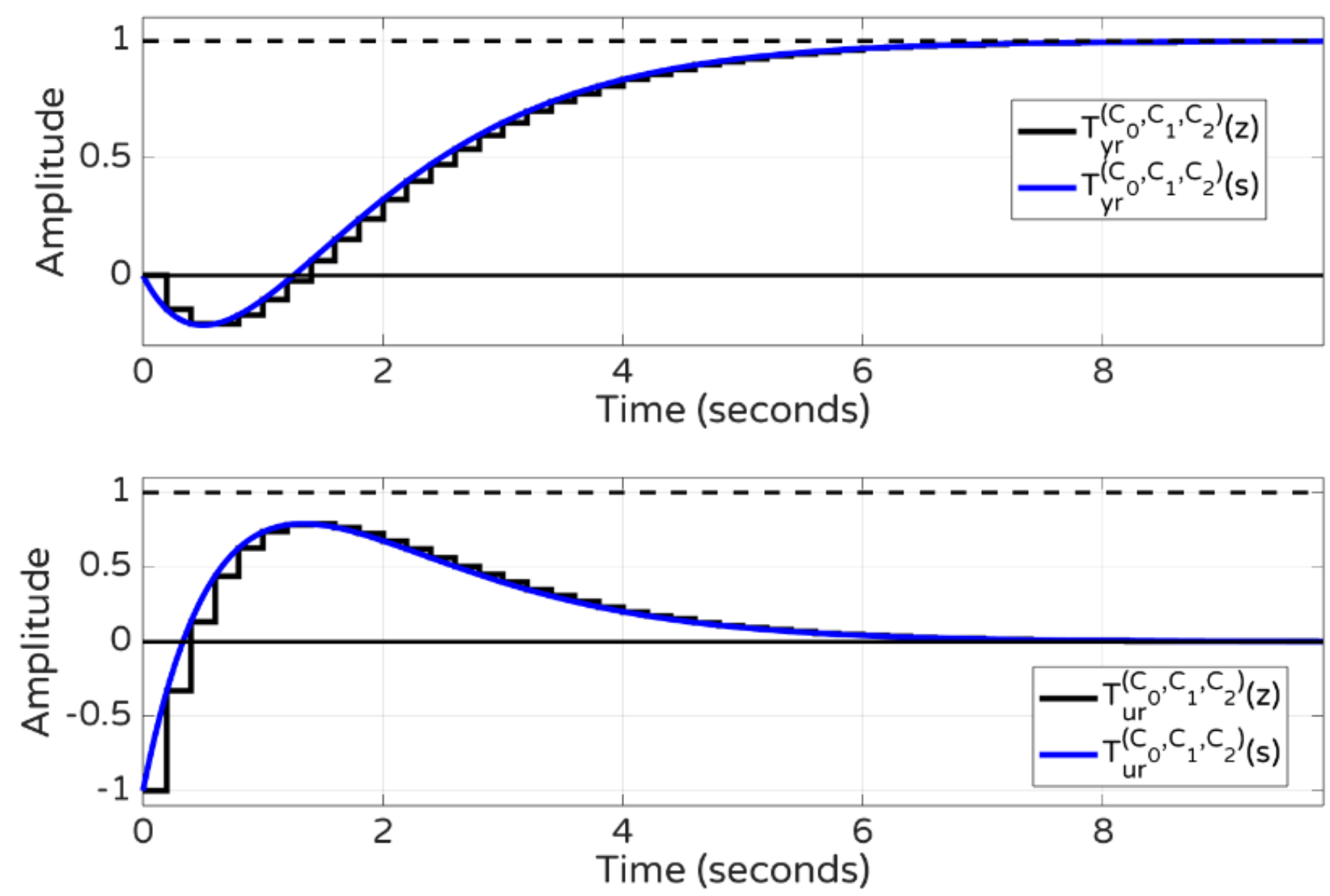

□

**Example 2**

A minimum-phase zero at $s=-1$ is added to the plant from Example 1.

$$P(s)=\frac{(s+1)(s-1)}{s(s-2)}$$

This plant satisfies the PIP because it is biproper— and thus has no zeros at infinity—and has only a single zero in the extended right half-plane $C_{+e}$. Since no pair of zeros exists in $C_{+e}$ to define an open interval $(z_i, z_{i+1})$ where the interlacing condition could fail, the requirement of an even number of poles is vacuously satisfied. Consequently, the plant admits a stable stabilizing controller.

The adopted coprime factorization of $P(s)$ is:

$$N(s)=\frac{s-1}{s+1},\ D(s)=\frac{s(s-2)}{(s+1)^2},\ X_1(s)=\frac{5s-1}{s+1},\ X_2(s)=-4$$

A model matching design is carried out as in Example 1 for the following cases:

**Case 1)** Without cancellation of the minimum-phase zero of $P(s)$ at $s=-1$.

Desired input-output closed loop model: $P_m^1(s)=-\frac{(s+1)(s-1)}{(s+1)^2}$

$C_{ff}(s)$, $C_{fb}(s)$, and the free $Q_r(s)$, $Q_y(s)$ are obtained:

$$C_{ff}(s)=0.25,\ C_{fb}(s)=\frac{-1.25s+0.25}{s+1},\ Q_r=-1,\ Q_y=0.$$

Note that since no LHP zero of the plant $P(s)$ is canceled, $Q_r(s)$ is bistable. Furthermore, since $C_{ff}$ and $C_{fb}$ are stable and the implementation $(C_{ff}, C_{fb})$ is acyclic, the controller is stable. The blocks of all the input-output feedback implementations are:

| Implementation | Blocks | | |
|---|---|---|---|
| $(C_0, C_1, C_2)$ | $C_0(s)=-1$ | $C_1(s)=-5$ | $C_2(s)=\frac{5s-1}{s+1}$ |
| $(\hat{C}_0, \hat{C}_1, \hat{C}_2)$ | $\hat{C}_0(s)=-1$ | $\hat{C}_1(s)=-5$ | $\hat{C}_2(s)=\frac{6s}{s+1}$ |
| $(\bar{C}_0, \bar{C}_1, \bar{C}_2)$ | $\bar{C}_0(s)=-1$ | $\bar{C}_1(s)=5$ | $\bar{C}_2(s)=\frac{-5s+1}{s+1}$ |
| $(\breve{C}_0, \breve{C}_1, \breve{C}_2)$ | $\breve{C}_0(s)=-1$ | $\breve{C}_1(s)=3$ | $\breve{C}_2(s)=\frac{5s-1}{s+1}$ |
| $(\tilde{C}_1, \tilde{C}_2)$ | ———— | $\tilde{C}_1(s)=3$ | $\tilde{C}_2(s)=\frac{-5s+1}{s+1}$ |

**Case 2)** The zero at $s=-1$ is canceled and replaced with a faster zero at $s=-5$.

Desired input-output closed loop model: $P_m^2(s) = -\dfrac{(s+5)(s-1)}{5(s+1)^2}$

$C_{ff}(s)$, $C_{fb}(s)$, and the free $Q_r(s)$, $Q_y(s)$ are obtained.

$$\begin{aligned} C_{ff}(s) &= \frac{0.05(s+5)}{(s+1)} \\ C_{fb}(s) &= \frac{-1.25s+0.25}{(s+1)} \end{aligned} \quad \Leftrightarrow \quad \begin{aligned} Q_r &= \frac{-0.2\ (s+5)}{(s+1)} \\ Q_y &= 0 \end{aligned}$$

Note that $Q_r(s)$ is bistable because the LHP zero of the plant $P(s)$ at $s=-1$ is canceled and replaced with the zero at $s=-5$.

The blocks of all the input-output feedback implementations are:

| Implementation | Blocks | | |
|---|---|---|---|
| $(C_0, C_1, C_2)$ | $C_0(s) = \dfrac{-0.2(s+5)}{(s+1)}$ | $C_1(s) = -5$ | $C_2(s) = \dfrac{5s-1}{s+1}$ |
| $(\hat{C}_0, \hat{C}_1, \hat{C}_2)$ | $\hat{C}_0(s) = \dfrac{-0.2(s+5)}{(s+1)}$ | $\hat{C}_1(s) = -5$ | $\hat{C}_2(s) = \dfrac{5.2s}{s+1}$ |
| $(\bar{C}_0, \bar{C}_1, \bar{C}_2)$ | $\bar{C}_0(s) = \dfrac{-0.2(s+5)}{(s+1)}$ | $\bar{C}_1(s) = \dfrac{25(s+1)}{(s+5)}$ | $\bar{C}_2(s) = \dfrac{-25(s-0.2)}{(s+5)}$ |
| $(\breve{C}_0, \breve{C}_1, \breve{C}_2)$ | $\breve{C}_0(s) = \dfrac{-5(s+1)}{(s+5)}$ | $\breve{C}_1(s) = \dfrac{-3.8s-3}{s+1}$ | $\breve{C}_2(s) = \dfrac{5s-1}{s+1}$ |
| $(\tilde{C}_1, \tilde{C}_2)$ | ———— | $\tilde{C}_1(s) = \dfrac{19s+15}{s+5}$ | $\tilde{C}_2(s) = \dfrac{-25s+5}{s+5}$ |

**Case 3)** The zero at $s=-1$ is canceled without replacement.

Desired input-output closed loop model: $P_m^3(s) = -\dfrac{(s-1)}{(s+1)^2}$

In this case $Q_r \notin \mathcal{U}$, because the relative degree of $P_m^3 = T_{yr}$ exceeds that of the plant $(\rho(P_m^3) > \rho(P))$. Consequently, the relative degree of $Q_r$ becomes greater than zero, rendering $Q_r^{-1}$ improper.

$\rho(P_m^3) = \rho(Q_r) + \rho(P) \Rightarrow \rho(Q_r) = \rho(P_m^3) - \rho(P) > 0 \Rightarrow Q_r^{-1} \notin \mathcal{R} \Rightarrow Q_r \notin \mathcal{U}$

$C_{ff}(s)$, $C_{fb}(s)$, $Q_r(s)$, and $Q_y(s)$ are obtained from the model matching design:

$$C_{ff}(s) = \frac{0.25}{s+1} \qquad \Leftrightarrow \qquad \boxed{Q_r = \frac{-1}{s+1} \notin \mathcal{U}}$$

$$C_{fb}(s) = \frac{-1.25s + 0.25}{(s+1)} \qquad \qquad Q_y = 0$$

The following table summarized the parametrization of all the input-output feedback implementations. It can be observed that the implementations $(\bar{C}_0, \bar{C}_1, \bar{C}_2)$, $(\breve{C}_0, \breve{C}_1, \breve{C}_2)$, and $(\tilde{C}_1, \tilde{C}_2)$ contain improper blocks, making them infeasible.

| Implementation | Blocks | | |
|---|---|---|---|
| $(C_0, C_1, C_2)$ | $C_0(s) = \frac{-1}{s+1}$ | $C_1(s) = -5$ | $C_2(s) = \frac{5s-1}{s+1}$ |
| $(\hat{C}_0, \hat{C}_1, \hat{C}_2)$ | $\hat{C}_0(s) = \frac{-1}{s+1}$ | $\hat{C}_1(s) = -5$ | $\hat{C}_2(s) = \frac{5s}{s+1}$ |
| $(\bar{C}_0, \bar{C}_1, \bar{C}_2)$ | $\bar{C}_0(s) = \frac{-1}{s+1}$ | $\bar{C}_1(s) = 5(s+1)$ | $\bar{C}_2(s) = 5s - 1$ |
| $(\breve{C}_0, \breve{C}_1, \breve{C}_2)$ | $\breve{C}_0(s) = -(s+1)$ | $\breve{C}_1(s) = \frac{-4s-3}{s+1}$ | $\breve{C}_2(s) = \frac{5s-1}{s+1}$ |
| $(\tilde{C}_1, \tilde{C}_2)$ | ——— | $\tilde{C}_1(s) = 4s + 3$ | $\tilde{C}_2(s) = -5s + 1$ |

□

**B. MIMO Case**

The control law $u = C_{ff} r - C_{fb} y \in \mathbb{R}^n$ operates with signals $r, y \in \mathbb{R}^m$. The dimensions of the matrices involved in the analysis of MIMO systems with $m$ outputs and $n$ inputs are as follows (2):

$P(s) \in M(\mathcal{R})^{m\times n}$, $N, \tilde{N} \in M(\mathcal{S})^{m\times n}$, $X_1, \tilde{X}_1 \in M(\mathcal{S})^{n\times m}$, $D, X_2 \in M(\mathcal{S})^{n\times n}$, $\tilde{D}, \tilde{X}_2 \in M(\mathcal{S})^{m\times m}$

$C_{ff}, C_{fb} \in M(\mathcal{R})^{n\times m}$ and $Q_r, Q_y \in M(\mathcal{R})^{n\times m}$.

For square MIMO systems (where $n = m$), all the aforementioned matrices are square with dimensions $n \times n$.

**B.1 Square MIMO Systems ($P(s) \in M(\mathcal{R})^{n\times n}$)**

Although the algebraic complexity increases due to the non-commutative nature of matrices, the fundamental principle holds. For the square MIMO case, the conclusion remains identical to the SISO case: the stability of all transfer function matrices between external inputs and outputs is equivalent to the internal stability of the system, provided there are no unstable pole-zero cancellations in the interconnections.

For the realizations $(\bar{C}_0,\bar{C}_1,\bar{C}_2)$, $(\breve{C}_0,\breve{C}_1,\breve{C}_2)$, and $(\tilde{C}_1,\tilde{C}_2)$, the parameter $Q_r$ is restricted to the set of units of $M(\mathcal{S})$, i.e., $Q_r \in \mathcal{U}(\mathcal{S}) : Q_r, Q_r^{-1} \in M(\mathcal{S})$.

The closed-loop transfer functions $T_{yr}^{(\alpha,\beta,\gamma)}(P)$ of the five MIMO input-output feedback implementations are given by Table I. All implementations are input-output equivalent. Furthermore, by Property 1 and under the parametrization of Fig. 15, all closed-loop transfer functions are identical to (24) and (25).

$$\begin{aligned} T_{yr}^{(C_0,C_1,C_2)}(P) &= T_{yr}^{(\hat{C}_0,\hat{C}_1,\hat{C}_2)}(P) = T_{yr}^{(\bar{C}_0,\bar{C}_1,\bar{C}_2)}(P) = T_{yr}^{(\tilde{C}_1,\tilde{C}_2)}(P) = T_{yr}^{(C_{ff},C_{fb})}(P) \\ &= T_{yr}^{(C_r,C_e,C_y)}(P) = NQ_r \end{aligned} \tag{53}$$

i) $$T_{yr}^{(C_0,C_1,C_2)}(P) = P(I + C_1 + C_2P)^{-1}C_0 = \underbrace{\tilde{D}^{-1}\tilde{N}}_{P}\,(I + \underbrace{X_2 - Q_y\tilde{N} - I}_{C_1} + \underbrace{(X_1 + Q_y\tilde{D})}_{C_2}\underbrace{\tilde{D}^{-1}\tilde{N}}_{P})^{-1}\underbrace{Q_r}_{C_0}$$
$$= \tilde{D}^{-1}\tilde{N}(X_2 + X_1\tilde{D}^{-1}\tilde{N})^{-1}Q_r = \left(Q_r^{-1}(X_2 + X_1\tilde{D}^{-1}\tilde{N})\tilde{N}^{-1}\tilde{D}\right)^{-1} = \Big(Q_r^{-1}\underbrace{(X_2\underbrace{DN^{-1}}_{\tilde{N}^{-1}\tilde{D}} + X_1)}_{N^{-1}}\Big)^{-1} = NQ_r$$

ii) $$\begin{aligned} T_{yr}^{(\hat{C}_0,\hat{C}_1,\hat{C}_2)}(P) &= P(I + \hat{C}_1 + \hat{C}_2P + \hat{C}_0P)^{-1}\hat{C}_0 \\ &= \underbrace{\tilde{D}^{-1}\tilde{N}}_{P}\,(I + \underbrace{(X_2 - Q_y\tilde{N} - I)}_{\hat{C}_1} + \underbrace{(X_1 + Q_y\tilde{D} - Q_r)}_{\hat{C}_2}\underbrace{\tilde{D}^{-1}\tilde{N}}_{P} + \underset{\hat{C}_0}{Q_r}\,\underbrace{\tilde{D}^{-1}\tilde{N}}_{P})^{-1}\underbrace{Q_r}_{\hat{C}_0} \\ &= \tilde{D}^{-1}\tilde{N}(X_2 + X_1\tilde{D}^{-1}\tilde{N})^{-1}Q_r = \Big(Q_r^{-1}\underbrace{(X_1N + X_2D)}_{I}N^{-1}\Big)^{-1} = (Q_r^{-1}N^{-1})^{-1} = NQ_r \end{aligned}$$

iii) $$\begin{aligned} T_{yr}^{(\bar{C}_0,\bar{C}_1,\bar{C}_2)}(P) &= P(I + \bar{C}_0\bar{C}_1 + \bar{C}_0\bar{C}_2P)^{-1}\bar{C}_0 = P(I + \underset{\bar{C}_0}{Q_r}\underbrace{Q_r^{-1}C_1}_{\bar{C}_1} + \underset{\bar{C}_0}{Q_r}\underbrace{Q_r^{-1}C_2}_{\bar{C}_2}P)^{-1}\underbrace{C_0^{-1}}_{\bar{C}_0} \\ &= P(I + C_1 + C_2P)^{-1}C_0 = T_{yr}^{(C_0,C_1,C_2)}(P) = NQ_r \end{aligned}$$

iv) $$\begin{aligned} T_{yr}^{(\breve{C}_0,\breve{C}_1,\breve{C}_2)}(P) &= P(I + \breve{C}_0\breve{C}_1 + \breve{C}_0\breve{C}_2P)^{-1} \\ &= ND^{-1}(I + \underbrace{Q_r^{-1}}_{\breve{C}_0}\underbrace{(X_2 - Q_y\tilde{N} - Q_r)}_{\breve{C}_1} + \underbrace{Q_r^{-1}}_{\breve{C}_0}\underbrace{(X_1 + Q_y\tilde{D})}_{\breve{C}_2}\underbrace{\tilde{D}^{-1}\tilde{N}}_{P})^{-1} \\ &= ND^{-1}(Q_r^{-1}D^{-1})^{-1} = ND^{-1}DQ_r = NQ_r \end{aligned}$$

v) $$\begin{aligned} T_{yr}^{(\tilde{C}_1,\tilde{C}_2)}(P) &= P(1 + \tilde{C}_1 + \tilde{C}_2P)^{-1} = ND^{-1}(I + \underbrace{Q_r^{-1}(X_2 - Q_y\tilde{N} - Q_r)}_{\tilde{C}_1} + \underbrace{Q_r^{-1}(X_1 + Q_y\tilde{D})}_{\tilde{C}_2}\underbrace{\tilde{D}^{-1}\tilde{N}}_{P})^{-1} \\ &= ND^{-1}(Q_r^{-1}D^{-1})^{-1} = ND^{-1}DQ_r = NQ_r \end{aligned}$$

Naturally, equalities (53) lead to the parametrizations shown in Fig. 15 by applying Property 1 to implementations $(C_{ff},C_{fb})$, $(C_0,C_1,C_2)$, $(\hat{C}_0,\hat{C}_1,\hat{C}_2)$, $(\bar{C}_0,\bar{C}_1,\bar{C}_2)$, $(\breve{C}_0,\breve{C}_1,\breve{C}_2)$, and $(\tilde{C}_1,\tilde{C}_2)$.

i) $(C_{ff}, C_{fb}) \Leftrightarrow (C_0, C_1, C_2)$

$$T_{yr}^{(C_{ff},C_{fb})}(P) = T_{yr}^{(C_0,C_1,C_2)}(P) \rightarrow (I + C_{fb}P)^{-1}C_{ff} = (I + C_1 + C_2P)^{-1}C_0$$

$$\underbrace{(C_0{}^{-1}(I + C_1) - C_{ff}{}^{-1})}_{0} + \underbrace{(C_0{}^{-1}C_2 - C_{ff}{}^{-1}C_{fb})}_{0}P = 0,$$

selecting $C_0 = Q_r$

$$C_{ff} = (I + C_1)^{-1}C_0 = (X_2 - Q_y\tilde{N})^{-1}Q_r$$

$$C_1 = \underbrace{C_0}_{Q_r}\ \underbrace{C_{ff}{}^{-1}}_{Q_r{}^{-1}(X_2 - Q_y\tilde{N})} - I = Q_rQ_r{}^{-1}(X_2 - Q_y\tilde{N}) - I = X_2 - Q_y\tilde{N} - I$$

$$C_{fb} = (I + C_1)^{-1}C_2 = (X_2 - Q_y\tilde{N})^{-1}(X_1 + Q_y\tilde{D})$$

$$C_2 = \underbrace{C_0}_{Q_r}\ \underbrace{C_{ff}{}^{-1}}_{Q_r{}^{-1}(X_2 - Q_y\tilde{N})}\ \underbrace{C_{fb}}_{(X_2 - Q_y\tilde{N})^{-1}(X_1 + Q_y\tilde{D})} = X_1 + Q_y\tilde{D}$$

$$\boxed{\begin{array}{c} (C_{ff}, C_{fb}) \Leftrightarrow (C_0, C_1, C_2) \\ \begin{array}{l} C_{ff} = (I + C_1)^{-1}C_0 = (X_2 - Q_y\tilde{N})^{-1}Q_r \\ C_{fb} = (I + C_1)^{-1}C_2 = (X_2 - Q_y\tilde{N})^{-1}(X_1 + Q_y\tilde{D}) \end{array} \Rightarrow \begin{array}{l} C_0 = Q_r \\ C_1 = X_2 - Q_y\tilde{N} - I \\ C_2 = X_1 + Q_y\tilde{D} \end{array} \end{array}} \quad (54)$$

ii) $(C_{ff}, C_{fb}) \Leftrightarrow (\hat{C}_0, \hat{C}_1, \hat{C}_2)$

$$T_{yr}^{(C_{ff},C_{fb})}(P) = T_{yr}^{(\hat{C}_0,\hat{C}_1,\hat{C}_2)}(P) \rightarrow (I + C_{fb}P)^{-1}C_{ff} = (1 + \hat{C}_1 + \hat{C}_2P + \hat{C}_0P)^{-1}\hat{C}_0$$

$$\underbrace{(\hat{C}_0{}^{-1}(I + \hat{C}_1) - C_{ff}{}^{-1})}_{0} + \underbrace{(\hat{C}_0{}^{-1}\hat{C}_2 - C_{ff}{}^{-1}C_{fb} + I)}_{0}P = 0$$

$$\begin{array}{l} C_{ff} = (I + \hat{C}_1)^{-1}\hat{C}_0 = (X_2 - Q_y\tilde{N})^{-1}Q_r \\ C_{fb} = C_{ff}(\hat{C}_0{}^{-1}\hat{C}_2 + I) = (I + \hat{C}_1)^{-1}(\hat{C}_2 + \hat{C}_0) \end{array} \Leftrightarrow \begin{array}{l} \hat{C}_1 = \underbrace{\hat{C}_0}_{Q_r}\ \underbrace{C_{ff}{}^{-1}}_{Q_r{}^{-1}(X_2 - Q_y\tilde{N})} - I = X_2 - Q_y\tilde{N} - I \\ \hat{C}_2 = \hat{C}_0C_{ff}{}^{-1}C_{fb} - \hat{C}_0 = X_1 + Q_y\tilde{D} - Q_r \end{array}$$

$$\boxed{\begin{array}{c} (C_{ff}, C_{fb}) \Leftrightarrow (\hat{C}_0, \hat{C}_1, \hat{C}_2) \\ \begin{array}{l} C_{ff} = (I + \hat{C}_1)^{-1}\hat{C}_0 = (X_2 - Q_y\tilde{N})^{-1}Q_r \\ C_{fb} = C_{ff}(\hat{C}_0{}^{-1}\hat{C}_2 + I) = (I + \hat{C}_1)^{-1}(\hat{C}_2 + \hat{C}_0) \end{array} \Rightarrow \begin{array}{l} \hat{C}_0 = Q_r \\ \hat{C}_1 = X_2 - Q_y\tilde{N} - I \\ \hat{C}_2 = X_1 + Q_y\tilde{D} - Q_r \end{array} \end{array}} \quad (55)$$

iii) $(C_{ff}, C_{fb}) \Leftrightarrow (\bar{C}_0, \bar{C}_1, \bar{C}_2)$

$$T_{yr}^{(C_{ff},C_{fb})}(P) = T_{yr}^{(\bar{C}_0,\bar{C}_1,\bar{C}_2)}(P) \rightarrow (I + C_{fb}P)^{-1}C_{ff} = (I + \bar{C}_0\bar{C}_1 + \bar{C}_0\bar{C}_2P)^{-1}\bar{C}_0$$

$$\underbrace{(\bar{C}_0{}^{-1} + \bar{C}_0{}^{-1}\bar{C}_0\bar{C}_1 - C_{ff}{}^{-1})}_{0} + \underbrace{(\bar{C}_0{}^{-1}\bar{C}_0\bar{C}_2 - C_{ff}{}^{-1}C_{fb})}_{0}P = 0$$

$$C_{ff} = (I + \bar{C}_0\bar{C}_1)^{-1}\bar{C}_0 = (X_2 - Q_y\tilde{N})^{-1}Q_r,\; C_{fb} = (I + \bar{C}_0\bar{C}_1)^{-1}\bar{C}_0\bar{C}_2 = (X_2 - Q_y\tilde{N})^{-1}(X_1 + Q_y\tilde{D})$$

$$\bar{C}_1 = Q_r{}^{-1}(X_2 - Q_y\tilde{N} - I),\; \bar{C}_2 = Q_r{}^{-1}(X_1 + Q_y\tilde{D})$$

$$\boxed{\begin{array}{c} (C_{ff}, C_{fb}) \Leftrightarrow (\bar{C}_0, \bar{C}_1, \bar{C}_2) \\ \begin{array}{l} C_{ff} = (I + \bar{C}_0\bar{C}_1)^{-1}\bar{C}_0 = (X_2 - Q_y\tilde{N})^{-1}Q_r \\ C_{fb} = (I + \bar{C}_0\bar{C}_1)^{-1}\bar{C}_0\bar{C}_2 = (X_2 - Q_y\tilde{N})^{-1}(X_1 + Q_y\tilde{D}) \end{array} \Rightarrow \begin{array}{l} \bar{C}_0 = Q_r \\ \bar{C}_1 = Q_r{}^{-1}(X_2 - Q_y\tilde{N} - I) \\ \bar{C}_2 = Q_r{}^{-1}(X_1 + Q_y\tilde{D}) \end{array} \\ Q_r, Q_r{}^{-1}, Q_y \in M(\mathcal{S}) \end{array}} \quad (56)$$

iv) $(C_{ff}, C_{fb}) \Leftrightarrow (\tilde{C}_1, \tilde{C}_2)$

$$T_{yr}^{(C_{ff},C_{fb})}(P) = T_{yr}^{(\tilde{C}_1,\tilde{C}_2)}(P) \rightarrow (I + C_{fb}P)^{-1}C_{ff} = (I + \tilde{C}_1 + \tilde{C}_2P)^{-1}$$

$$\underbrace{(C_{ff}{}^{-1} - I - \tilde{C}_1)}_{0} + \underbrace{(C_{ff}{}^{-1}C_{fb} - \tilde{C}_2)}_{0}P = 0$$

$$C_{ff} = (I + \tilde{C}_1)^{-1} = (X_2 - Q_y\tilde{N})^{-1}Q_r$$

$$\tilde{C}_1 = C_{ff}{}^{-1} - I = Q_r{}^{-1}(X_2 - Q_y\tilde{N}) - I = Q_r{}^{-1}(X_2 - Q_y\tilde{N} - Q_r)$$

$$\tilde{C}_2 = C_{ff}{}^{-1}C_{fb} = Q_r{}^{-1}(X_2 - Q_y\tilde{N})(X_2 - Q_y\tilde{N})^{-1}(X_1 + Q_y\tilde{D}) = Q_r{}^{-1}(X_1 + Q_y\tilde{D})$$

$$C_{fb} = (I + \tilde{C}_1)^{-1}\tilde{C}_2 = (X_2 - Q_y\tilde{N})^{-1}(X_1 + Q_y\tilde{D})$$

$$\boxed{\begin{array}{c} (C_{ff}, C_{fb}) \Leftrightarrow (\tilde{C}_1, \tilde{C}_2) \\ \begin{array}{l} C_{ff} = C_{ff} = (I + \tilde{C}_1)^{-1} = (X_2 - Q_y\tilde{N})^{-1}Q_r \\ C_{fb} = (I + \tilde{C}_1)^{-1}\tilde{C}_2 = (X_2 - Q_y\tilde{N})^{-1}(X_1 + Q_y\tilde{D}) \end{array} \Leftrightarrow \begin{array}{l} \tilde{C}_1 = C_{ff}{}^{-1} - I = Q_r{}^{-1}(X_2 - Q_y\tilde{N} - Q_r) \\ \tilde{C}_2 = C_{ff}{}^{-1}C_{fb} = Q_r{}^{-1}(X_1 + Q_y\tilde{D}) \end{array} \\ Q_r, Q_r{}^{-1}, Q_y \in M(\mathcal{S}) \end{array}} \quad (57)$$

All implementations input-output feedback $(C_0, C_1, C_2)$, $(\hat{C}_0, \hat{C}_1, \hat{C}_2)$, $(\bar{C}_0, \bar{C}_1, \bar{C}_2)$, $(\breve{C}_0, \breve{C}_1, \breve{C}_2)$, and $(\tilde{C}_1, \tilde{C}_2)$ are input-output equivalent to the $(C_{ff}, C_{fb})$ implementation; consequently, they all share the same six transfer matrices (16). For instance, for implementation $(C_0, C_1, C_2)$, the gang of six is:

$$\begin{cases} T_{yr}^{(C_0,C_1,C_2)}(P) = P(I_n + C_1 + C_2P)^{-1}C_0 = NQ_r \\ T_{yd}^{(C_0,C_1,C_2)}(P) = P(I_n + C_1 + C_2P)^{-1}(I_n + C_1) = (\tilde{X}_2 - NQ_y)\tilde{N} \\ T_{yn}^{(C_0,C_1,C_2)}(P) = I_m - P(I_n + C_1 + C_2P)^{-1}C_2 = (\tilde{X}_2 - NQ_y)\tilde{D} \\ T_{ur}^{(C_0,C_1,C_2)}(P) = (I_n + C_1 + C_2P)^{-1}C_0 = DQ_r \\ T_{ud}^{(C_0,C_1,C_2)}(P) = -(I_n + C_1 + C_2P)^{-1}C_2P = -(\tilde{X}_1 + DQ_y)\tilde{N} \\ T_{un}^{(C_0,C_1,C_2)}(P) = -(I_n + C_1 + C_2P)^{-1}C_2 = -(\tilde{X}_1 + DQ_y)\tilde{D} \end{cases} \tag{58}$$

Table III and Figure 15 summarize the parametrization of the five I/O feedback implementations of the control law (5) for square MIMO systems.

| **Implementation** | **Blocks** | | |
|---|---|---|---|
| $(C_0, C_1, C_2)$ | $C_0 = Q_r$ | $C_1 = X_2 - Q_y\tilde{N} - I$ | $C_2 = X_1 + Q_y\tilde{D}$ |
| $(\hat{C}_0, \hat{C}_1, \hat{C}_2)$ | $\hat{C}_0 = Q_r$ | $\hat{C}_1 = X_2 - Q_y\tilde{N} - I$ | $\hat{C}_2 = X_1 + Q_y\tilde{D} - Q_r$ |
| $(\bar{C}_0, \bar{C}_1, \bar{C}_2)$ | $\bar{C}_0 = Q_r$ | $\bar{C}_1 = Q_r^{-1}(X_2 - Q_y\tilde{N} - I)$ | $\bar{C}_2 = Q_r^{-1}(X_1 + Q_y\tilde{D})$ |
| $(\breve{C}_0, \breve{C}_1, \breve{C}_2)$ | $\breve{C}_0 = Q_r^{-1}$ | $\breve{C}_1 = X_2 - Q_y\tilde{N} - Q_r$ | $\breve{C}_2 = X_1 + Q_y\tilde{D}$ |
| $(\tilde{C}_1, \tilde{C}_2)$ | ——— | $\tilde{C}_1 = Q_r^{-1}(X_2 - Q_y\tilde{N} - Q_r)$ | $\tilde{C}_2 = Q_r^{-1}(X_1 + Q_y\tilde{D})$ |

Table III: Parametrization of the five MIMO input-output feedback implementations for square MIMO systems. The parametrizations of the $(C_0, C_1, C_2)$ and $(\hat{C}_0, \hat{C}_1, \hat{C}_2)$ implementations hold for non-square MIMO systems

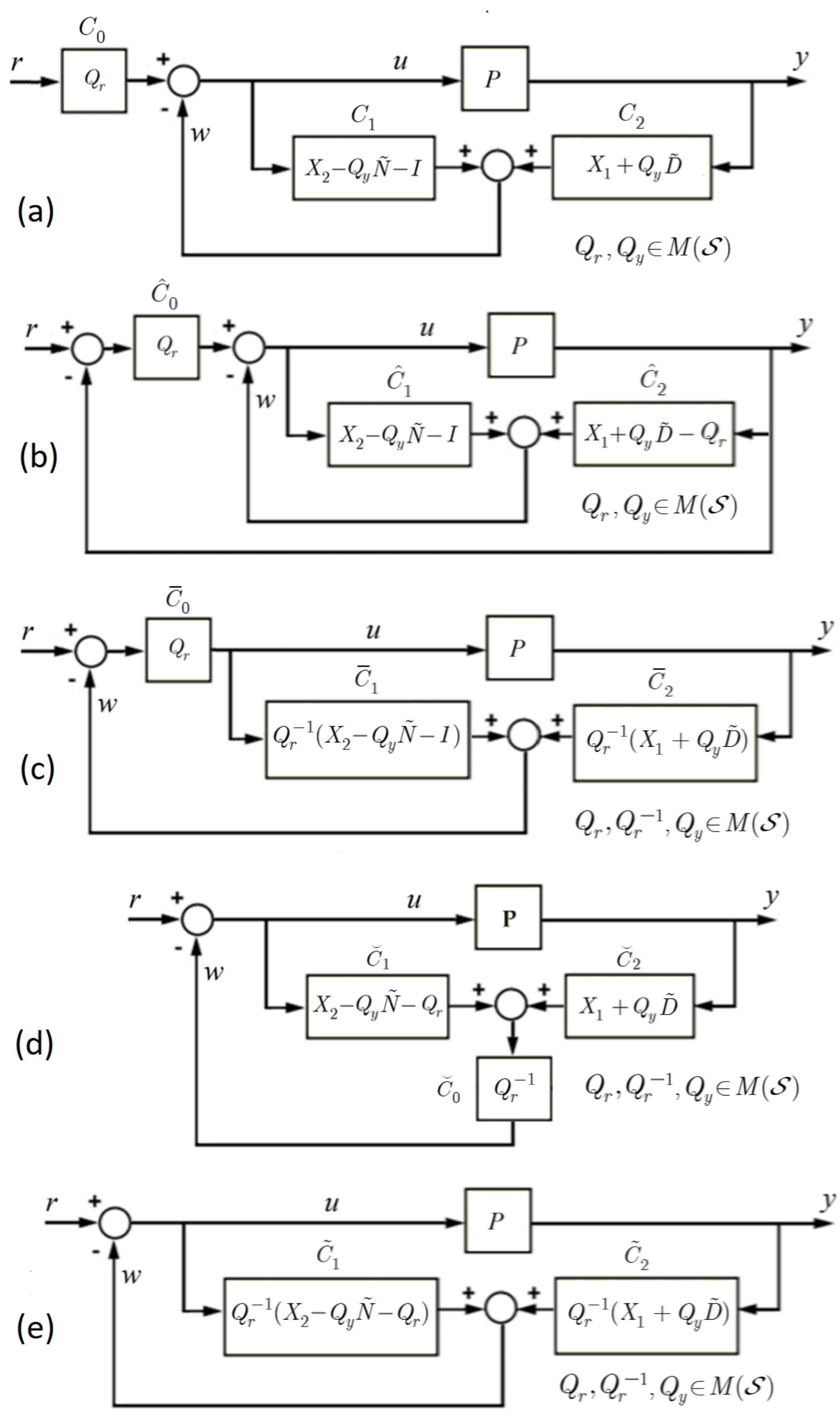

Fig. 15. Stabilization of square MIMO systems with input-output feedback implementations. (a) input-output feedback with prefilter $(C_0, C_1, C_2)$, (b) two-stage Input-Output feedback $(\hat{C}_0, \hat{C}_1, \hat{C}_2)$, (c) input-output feedback $(\bar{C}_0, \bar{C}_1, \bar{C}_2)$, (d) observer-controller $(\breve{C}_0, \breve{C}_1, \breve{C}_2)$, (e) two-block input-output feedback $(\tilde{C}_1, \tilde{C}_2)$.

### B.2 Non-square MIMO Systems ( $P(s) \in M(\mathcal{R})^{m\times n}, m \neq n$ )

In both cases:

i) More outputs than inputs ( $m > n$ )—systems with redundant sensors ("Tall" systems), where the plant matrix $P$ has "skinny column" dimensions.

ii) More inputs than outputs ( $m < n$ )— systems with redundant actuators ("Wide" systems), where the plant matrix $P$ is a "fat row."

Only implementations $(C_0, C_1, C_2)$ and $(\hat{C}_0, \hat{C}_1, \hat{C}_2)$ can be utilized. This is because the controller's free parameters $Q_r, Q_y \in M(\mathcal{S})^{n\times m}$ are non-square matrices; consequently, $Q_r^{-1}$ does not exist, rendering implementations $(\bar{C}_0, \bar{C}_1, \bar{C}_2)$, $(\breve{C}_0, \breve{C}_1, \breve{C}_2)$, and $(\tilde{C}_1, \tilde{C}_2)$ infeasible.

Figures 14(a) and (b), along with Table IV, summarize the parametrization of $(C_0, C_1, C_2)$ and $(\hat{C}_0, \hat{C}_1, \hat{C}_2)$ implementations for non-square MIMO systems.

| Implementation | Blocks | | |
|---|---|---|---|
| $(C_0, C_1, C_2)$ | $\underbrace{C_0}_{n\times m} = Q_r$ | $\underbrace{C_1}_{n\times n} = X_2 - Q_y\tilde{N} - I_n$ | $\underbrace{C_2}_{n\times m} = X_1 + Q_y\tilde{D}$ |
| $(\hat{C}_0, \hat{C}_1, \hat{C}_2)$ | $\underbrace{\hat{C}_0}_{n\times m} = Q_r$ | $\underbrace{\hat{C}_1}_{n\times n} = X_2 - Q_y\tilde{N} - I_n$ | $\underbrace{\hat{C}_2}_{n\times m} = X_1 + Q_y\tilde{D} - Q_r$ |

Table IV: Parametrization of the five MIMO input-output feedback implementations for non-square MIMO systems

**Application Examples for MIMO Systems**

**Example 3**

Consider the following square MIMO plant with two outputs and two inputs:

$$P(s) = \begin{bmatrix} \dfrac{(s-1)}{s(s-2)} & \dfrac{1}{s+1} \\ \dfrac{1}{s+2} & \dfrac{1}{s+3} \end{bmatrix}$$

The poles of the plant are given by $\{s, (s-2), (s+1), (s+2), (s+3)\}$, and the finite transmission zeros of the plant are given by $\{-5.372,\ 0.372\}$. The plant possesses two transmission zeros at infinity, with structural orders given by $k_1 = 1$ and $k_2 = 2$ (yielding a total infinite zero multiplicity of 3). Since the number of poles between the real transmission zeros at $0.372$ and $\infty$ is odd (namely 1), the plant does not satisfy the PIP.

The adopted doubly coprime factorization of $P(s)$ is:

$$D(s) = \tilde{D}(s) = \begin{bmatrix} \dfrac{s(s-2)}{(s+1)^2} & 0 \\ 0 & 1 \end{bmatrix}$$

$$N(s) = \begin{bmatrix} \dfrac{(s-1)}{(s+1)^2} & \dfrac{1}{s+1} \\ \dfrac{s(s-2)}{(s+1)^2(s+2)} & \dfrac{1}{s+3} \end{bmatrix}, \; \tilde{N}(s) = \begin{bmatrix} \dfrac{(s-1)}{(s+1)^2} & \dfrac{s(s-2)}{(s+1)^3} \\ \dfrac{1}{s+2} & \dfrac{1}{s+3} \end{bmatrix}$$

$$X_1(s) = \begin{bmatrix} \dfrac{14s-1}{s+1} & 0 \\ 0 & 0 \end{bmatrix}, \; X_2(s) = \begin{bmatrix} \dfrac{s-9}{s+1} & \dfrac{-14s+1}{(s+1)^2} \\ 0 & 1 \end{bmatrix}$$

$$\tilde{X}_1(s) = \begin{bmatrix} \dfrac{14s-1}{s+1} & 0 \\ 0 & 0 \end{bmatrix}, \; \tilde{X}_2(s) = \begin{bmatrix} \dfrac{s-9}{s+1} & 0 \\ \dfrac{-14s+1}{(s+1)(s+2)} & 1 \end{bmatrix}.$$

By choosing the following free parameters $Q_r$ and $Q_y$:

$$Q_r(s) = \begin{bmatrix} -\dfrac{(s+2)}{(s+1)} & 0 \\ 0 & -\dfrac{(s+2)}{(s+1)} \end{bmatrix}, \; Q_y(s) = \begin{bmatrix} 1 & 0 \\ 0 & 1 \end{bmatrix},$$

the parametrizations for the five I/O feedback implementations are obtained, which are presented in the table below. It can be observed that the blocks of all implementations are stable.

| Implementation | Blocks | | |
|---|---|---|---|
| $(C_0, C_1, C_2)$ | $C_0(s) = \begin{bmatrix} -\frac{(s+2)}{(s+1)} & 0 \\ 0 & -\frac{(s+2)}{(s+1)} \end{bmatrix}$ | $C_1(s) = \begin{bmatrix} -\frac{(11s+9)}{(s+1)^2} & -\frac{15(s-0.082)(s+0.82)}{(s+1)^3} \\ -\frac{1}{s+2} & -\frac{1}{(s+3)} \end{bmatrix}$ | $C_2(s) = \begin{bmatrix} \frac{15(s-0.082)(s+0.82)}{(s+1)^2} & 0 \\ 0 & 1 \end{bmatrix}$ |
| $(\hat{C}_0, \hat{C}_1, \hat{C}_2)$ | $\hat{C}_0(s) = \begin{bmatrix} -\frac{(s+2)}{(s+1)} & 0 \\ 0 & -\frac{(s+2)}{(s+1)} \end{bmatrix}$ | $\hat{C}_1(s) = \begin{bmatrix} -\frac{(11s+9)}{(s+1)^2} & -\frac{15(s-0.082)(s+0.82)}{(s+1)^3} \\ -\frac{1}{s+2} & -\frac{1}{(s+3)} \end{bmatrix}$ | $\hat{C}_2(s) = \begin{bmatrix} \frac{16\ (s+0.08)\ (s+0.8)}{(s+1)^2} & 0 \\ 0 & \frac{s+1.5}{s+1} \end{bmatrix}$ |
| $(\bar{C}_0, \bar{C}_1, \bar{C}_2)$ | $\bar{C}_0(s) = \begin{bmatrix} -\frac{(s+2)}{(s+1)} & 0 \\ 0 & -\frac{(s+2)}{(s+1)} \end{bmatrix}$ | $\bar{C}_1(s) = \begin{bmatrix} \frac{11(s+0.82)}{(s+1)(s+2)} & \frac{15(s-0.082)(s+0.82)}{(s+1)^2(s+2)} \\ \frac{(s+1)}{(s+2)^2} & \frac{s+1}{s+3} \end{bmatrix}$ | $\bar{C}_2(s) = \begin{bmatrix} -\frac{15(s+0.82)(s-0.082)}{(s+2)(s+1)} & 0 \\ 0 & -\frac{s+1}{s+2} \end{bmatrix}$ |
| $(\breve{C}_0, \breve{C}_1, \breve{C}_2)$ | $\breve{C}_0 = \begin{bmatrix} -\frac{(s+1)}{(s+2)} & 0 \\ 0 & -\frac{(s+1)}{(s+2)} \end{bmatrix}$ | $\breve{C}_1 = \begin{bmatrix} \frac{2(s-3.8)(s+0.8)}{(s+1)^2} & \frac{-1}{s+2} \\ \frac{-15(s-0.08)(s+0.8)}{(s+1)^3} & \frac{2(s+2)^2}{(s+1)(s+3)} \end{bmatrix}$ | $\breve{C}_2(s) = \begin{bmatrix} \frac{15(s-0.82)(s+0.082)}{(s+1)^2} & 0 \\ 0 & 1 \end{bmatrix}$ |
| $(\tilde{C}_1, \tilde{C}_2)$ | ——— | $\tilde{C}_1(s) = \begin{bmatrix} -\frac{2(s-3.8)(s+0.8)}{(s+1)(s+2)} & \frac{15(s-0.082)(s+0.82)}{(s+1)^2(s+2)} \\ \frac{s+1}{s+2} & -\frac{2(s+2)}{(s+3)} \end{bmatrix}$ | $\tilde{C}_2(s) = \begin{bmatrix} -\frac{15(s+0.82)(s-0.082)}{(s+2)(s+1)} & 0 \\ 0 & -\frac{(s+1)}{(s+2)} \end{bmatrix}$ |

The following figure illustrates the step response from the input $u = [u_1\ u_2]^{\mathrm{T}}$ to the output $y = [y_1\ y_2]^{\mathrm{T}}$ for the $(C_0, C_1, C_2)$ implementation. By (53), all other implementations yield the same result.

$$T_{yr}^{(C_0,C_1,C_2)}(P) = T_{yr}^{(\hat{C}_0,\hat{C}_1,\hat{C}_2)}(P) = T_{yr}^{(\bar{C}_0,\bar{C}_1,\bar{C}_2)}(P) = T_{yr}^{(\breve{C}_0,\breve{C}_1,\breve{C}_2)}(P) = T_{yr}^{(\tilde{C}_1,\tilde{C}_2)}(P) = NQ_r$$

$$= \begin{bmatrix} \dfrac{-(s-1)(s+2)}{(s+1)^3} & \dfrac{-(s+2)}{(s+1)^2} \\ \dfrac{-s(s-2)}{(s+1)^3} & \dfrac{-(s+2)}{(s+1)(s+3)} \end{bmatrix}$$

□

**Example 4**

Consider the following non-square MIMO plant with two outputs and one input ("Tall" system):

$$P(s) = \begin{bmatrix} \dfrac{(s-1)}{s(s-2)} \\ \dfrac{1}{s+1} \end{bmatrix}$$

The adopted doubly coprime factorization of $P(s)$ is:

$$D(s) = \frac{s(s-2)}{(s+1)^2},\ \tilde{D}(s) = \begin{bmatrix} \dfrac{s(s-2)}{(s+1)^2} & 0 \\ 0 & 1 \end{bmatrix},\ N(s) = \begin{bmatrix} \dfrac{(s-1)}{(s+1)^2} \\ \dfrac{s(s-2)}{(s+1)^3} \end{bmatrix},\ \tilde{N}(s) = \begin{bmatrix} \dfrac{(s-1)}{(s+1)^2} \\ \dfrac{1}{s+1} \end{bmatrix}$$

$$X_1(s) = \begin{bmatrix} \dfrac{41s-1}{(s+1)^2} & 0 \end{bmatrix},\ X_2(s) = \frac{(s-2.66)\ (s+8.66)}{(s+1)^2}$$

$$\tilde{X}_1(s) = \begin{bmatrix} \frac{41s-1}{(s+1)^2} & 0 \end{bmatrix},\ \tilde{X}_2(s) = \begin{bmatrix} \frac{(s-2.66)\ (s+8.66)}{(s+1)^2} & 0 \\ \frac{-41s+1}{(s+1)^3} & 1 \end{bmatrix}.$$

By choosing the following free parameters $Q_r$ and $Q_y$:

$$Q_r(s) = \begin{bmatrix} -1 & 0 \end{bmatrix},\ Q_y(s) = \begin{bmatrix} 1 & 0 \end{bmatrix}$$

the parametrization of $(C_0, C_1, C_2)$ and $(\hat{C}_0, \hat{C}_1, \hat{C}_2)$ implementations for non-square MIMO systems are obtained, which are presented in the table below.

| Implementation | Blocks | | |
|---|---|---|---|
| $(C_0, C_1, C_2)$ | $C_0 = \begin{bmatrix} -1 & 0 \end{bmatrix}$ | $C_1 = \frac{3s-23}{(s+1)^2}$ | $C_2 = \begin{bmatrix} \frac{(s+39)\ (s-0.026)}{(s+1)^2} & 0 \end{bmatrix}$ |
| $(\hat{C}_0, \hat{C}_1, \hat{C}_2)$ | $\hat{C}_0 = \begin{bmatrix} -1 & 0 \end{bmatrix}$ | $\hat{C}_1 = \frac{3s-23}{(s+1)^2}$ | $\hat{C}_2 = \begin{bmatrix} \frac{2s(s+20.5)}{(s+1)^2} & 0 \end{bmatrix}$ |

The following figure illustrates the step response from the reference input $r_1$ to the output $y = [y_1\ y_2]^{\mathrm{T}}$ for the $(C_0, C_1, C_2)$ implementation. By (53), the $(\hat{C}_0, \hat{C}_1, \hat{C}_2)$ implementation yield the same result.

$$\underbrace{T_{yr}^{(C_0,C_1,C_2)}(P)}_{2\times2} = \underset{2\times1}{N}\ \underset{1\times2}{Q_r} = \begin{bmatrix} N(1)\underset{-1}{Q_r(1)} & N(1)\underset{0}{Q_r(2)} \\ N(2)\underset{-1}{Q_r(1)} & N(2)\underset{0}{Q_r(2)} \end{bmatrix} = \begin{bmatrix} -N(1) & 0 \\ -N(2) & 0 \end{bmatrix} = \begin{bmatrix} -\frac{(s-1)}{(s+1)^2} & 0 \\ -\frac{s(s-2)}{(s+1)^3} & 0 \end{bmatrix}$$

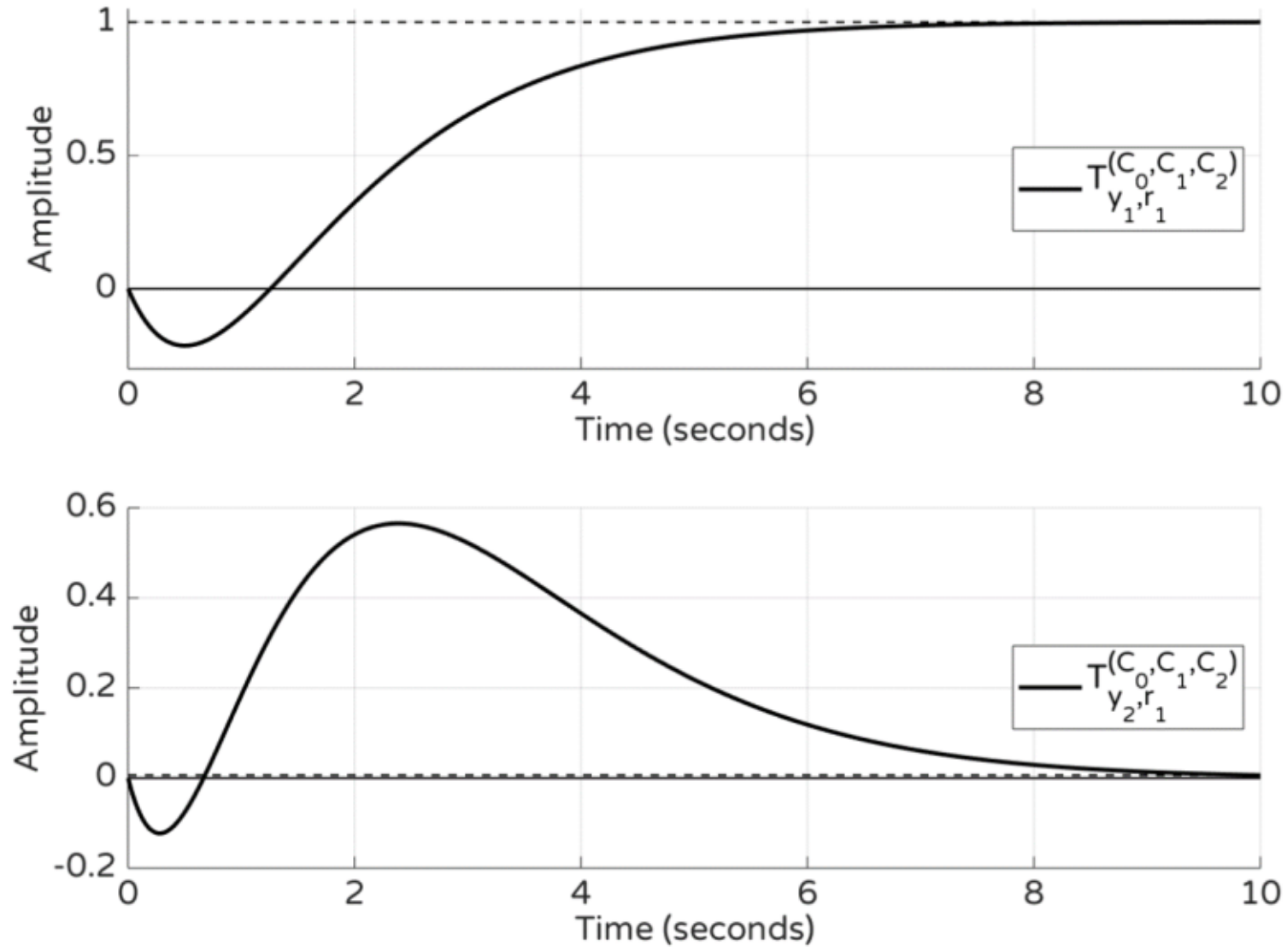

□

**Example 5**

Consider the following non-square MIMO plant with one output and two inputs ("Wide" system):

$$P(s) = \begin{bmatrix} \dfrac{(s-1)}{s(s-2)} & \dfrac{1}{s+1} \end{bmatrix}$$

The adopted doubly coprime factorization of $P(s)$ is:

$$D(s) = \begin{bmatrix} \dfrac{s(s-2)}{(s+1)^2} & 0 \\ 0 & 1 \end{bmatrix},\ \tilde{D}(s) = \frac{s(s-2)}{(s+1)^2},\ N(s) = \begin{bmatrix} \dfrac{(s-1)}{(s+1)^2} & \dfrac{1}{s+1} \end{bmatrix},\ \tilde{N}(s) = \frac{s(s-2)}{(s+1)^3}$$

$$X_1(s) = \begin{bmatrix} \dfrac{41s-1}{(s+1)^2} & 0 \end{bmatrix},\ X_2(s) = \begin{bmatrix} \dfrac{(s-2.66)\ (s+8.66)}{(s+1)^2} & \dfrac{-41s+1}{(s+1)^3} \\ 0 & 0 \end{bmatrix}$$

$$\tilde{X}_1(s) = \begin{bmatrix} \dfrac{41s-1}{(s+1)^2} \\ 0 \end{bmatrix},\ \tilde{X}_2(s) = \frac{(s-2.66)\ (s+8.66)}{(s+1)^2}$$

By choosing the following free parameters $Q_r$ and $Q_y$:

$$Q_r(s) = \begin{bmatrix} -1 \\ 0 \end{bmatrix},\ Q_y(s) = \begin{bmatrix} 1 \\ 0 \end{bmatrix}$$

The parametrization of $(C_0, C_1, C_2)$ and $(\hat{C}_0, \hat{C}_1, \hat{C}_2)$ implementations for non-square MIMO systems are obtained, which are presented in the table below.

| Implementation | Blocks | | |
|---|---|---|---|
| $(C_0, C_1, C_2)$ | $C_0 = \begin{bmatrix} -1 \\ 0 \end{bmatrix}$ | $C_1 = \begin{bmatrix} \dfrac{3s-23}{(s+1)^2} & \dfrac{-(s+39)(s-0.026)}{(s+1)^3} \\ 0 & 0 \end{bmatrix}$ | $C_2 = \begin{bmatrix} \dfrac{(s+39)\ (s-0.026)}{(s+1)^2} \\ 0 \end{bmatrix}$ |
| $(\hat{C}_0, \hat{C}_1, \hat{C}_2)$ | $\hat{C}_0 = \begin{bmatrix} -1 \\ 0 \end{bmatrix}$ | $\hat{C}_1 = \begin{bmatrix} \dfrac{3s-23}{(s+1)^2} & \dfrac{-(s+39)(s-0.026)}{(s+1)^3} \\ 0 & 0 \end{bmatrix}$ | $\hat{C}_2 = \begin{bmatrix} \dfrac{2s(s+20.5)}{(s+1)^2} \\ 0 \end{bmatrix}$ |

The following figure illustrates the step response from the scalar input reference input $r$ to the scalar output $y$ for the $(C_0, C_1, C_2)$ implementation. By (53), the $(\hat{C}_0, \hat{C}_1, \hat{C}_2)$ implementation yield the same result:

$$T_{yr}^{(C_0, C_1, C_2)}(P) = N\,Q_r = N(1)Q_r(1) + N(2)Q_r(2) = -N(1) = -\frac{(s-1)}{(s+1)^2}$$

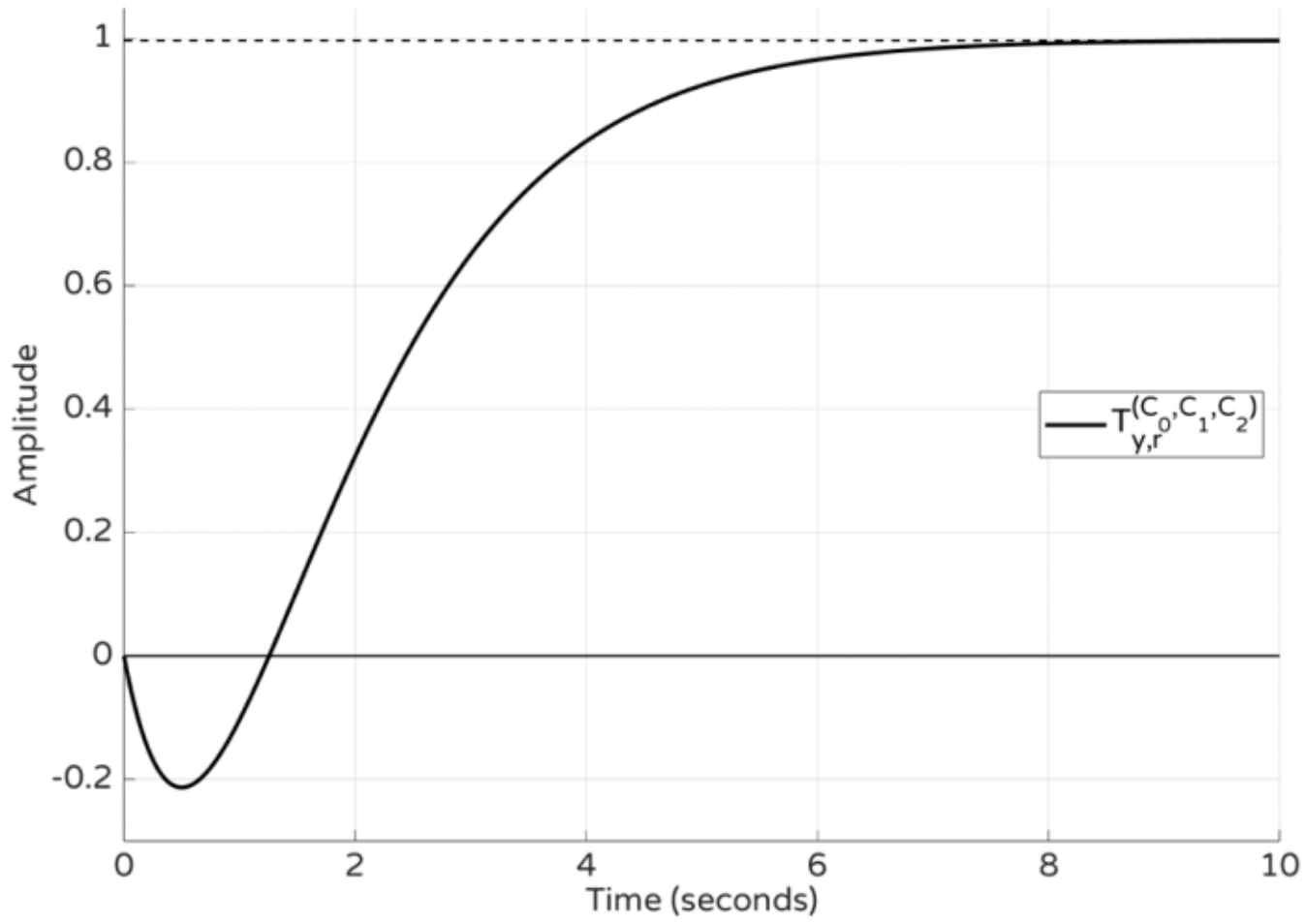

□

## IV. CONCLUSIONS

In this paper, we have presented an implementation method for two-parameter stabilizing controllers using only stable blocks, which is universally valid for all regular SISO and MIMO plants, as well as for both continuous- and discrete-time two-parameter controllers. The method is based on the input-output feedback control structures shown in Fig. 14 for the SISO case and Fig. 15 for the MIMO case, alongside their respective Youla-Kučera-based parametrizations presented in Tables II and III. From a practical standpoint, this framework is particularly valuable for implementing inherently unstable controllers, where the use of stable blocks offers significant advantages. Theoretically, by virtue of the direct correspondence between input-output feedback structures and observer-based state-space control, this work also holds potential theoretical significance.